\documentclass[aps,twocolumn,
superscriptaddress,
footinbib,
prb,floatfix]{revtex4-2}

\usepackage{graphicx} 
\usepackage{multirow}
\usepackage{graphicx}
\usepackage{comment}
\usepackage{enumitem}
\usepackage{mwe}
\usepackage{mathtools}
\usepackage{bbold}
\usepackage{amsmath,amssymb,bm}
\usepackage{mathrsfs}
\usepackage{MnSymbol}
\usepackage{graphicx}
\usepackage{epstopdf}
\usepackage[normalem]{ulem} 
\usepackage[caption=false]{subfig}
\usepackage{comment}
\usepackage{subfig}
\usepackage[usenames,dvipsnames]{color}
\usepackage{hyperref}
\usepackage[nameinlink]{cleveref}
\usepackage{natbib}
\usepackage{xcolor, colortbl}
\usepackage{physics}
\usepackage{multirow}
\usepackage{stackengine}
\usepackage{dsfont}
\usepackage{relsize}
\usepackage{framed}
\usepackage[utf8]{inputenc} 
\usepackage[acronym]{glossaries}
\usepackage{ragged2e}
\usepackage{algorithm}
\usepackage{algpseudocode} 
\usepackage{amsthm} 
\usepackage{dsfont}
\usepackage[utf8]{inputenc}

\begin{document}
\title{Benchmarking the quality of multiplexed qubit readout beyond assignment fidelity}
\author{Andras Di Giovanni}
\affiliation{Physikalisches Institut, Karlsruher Institut für Technologie, Kaiserstraße 12, 76131 Karlsruhe, Germany}
\author{Adrian Skasberg Aasen}
\affiliation{Kirchhoff-Institut f\"{u}r Physik, Universit\"{a}t Heidelberg, Im Neuenheimer Feld 227, 69120 Heidelberg, Germany}
\affiliation{Institut für Festk\"{o}rpertheorie und -optik,  Friedrich-Schiller-Universit\"{a}t Jena, Max-Wien-Platz 1, 07743 Jena, Germany}
\author{Jürgen Lisenfeld}
\affiliation{Physikalisches Institut, Karlsruher Institut für Technologie, Kaiserstraße 12, 76131 Karlsruhe, Germany}
\author{Martin Gärttner}
\affiliation{Institut für Festk\"{o}rpertheorie und -optik,  Friedrich-Schiller-Universit\"{a}t Jena, Max-Wien-Platz 1, 07743 Jena, Germany}
\author{Hannes Rotzinger}
\affiliation{Physikalisches Institut, Karlsruher Institut für Technologie, Kaiserstraße 12, 76131 Karlsruhe, Germany}
\affiliation{Institut für QuantenMaterialien und Technologien, Karlsruher Institut für Technologie, Kaiserstraße 12, 76131 Karlsruhe, Germany}
\author{Alexey V. Ustinov}
\affiliation{Physikalisches Institut, Karlsruher Institut für Technologie, Kaiserstraße 12, 76131 Karlsruhe, Germany}
\affiliation{Institut für QuantenMaterialien und Technologien, Karlsruher Institut für Technologie, Kaiserstraße 12, 76131 Karlsruhe, Germany}
\date{\today}


\begin{abstract}
The accurate measurement of quantum two-level objects (qubits) is crucial for developing quantum computers. Over the last decade, the measure of choice for benchmarking readout routines for superconducting qubits has been assignment fidelity. However, this method only focuses on the preparation of computational basis states and therefore does not provide a complete characterization of the readout.
Here, we expand the focus to the use of detector tomography to fully characterize multiqubit readout of superconducting transmon qubits.
The impact of different readout parameters on the rate of information extraction is studied using quantum state reconstruction infidelity as a proxy. The results are then compared with assignment fidelities, showing good agreement for separable two-qubit states. We therefore propose the rate of infidelity convergence as a validation tool for assignment fidelity and a more comprehensive benchmark for single- and multiqubit readout optimization. 
To make the best use of experimental
resources, we investigate the most efficient distribution of a limited shot budget between detector tomography and state reconstruction within the context of single- and two-qubit experiments. To address the growing interest in three-qubit gates and test scalability of the validation tool, we perform three-qubit quantum state tomography that goes beyond conventional readout-error-mitigation methods and find a factor of 30 reduction in quantum infidelity. Our results demonstrate that qubit readout correlations are not induced by a significantly reduced state distinguishability. Consequently, correlation coefficients can serve as a valuable tool in qubit readout optimization.
\end{abstract}

\maketitle


\section{Introduction}
State-of-the-art quantum computing experiments have started pushing the boundary to fault tolerance \cite{Google2023, Google2024}. Although gate-error-mitigation approaches such as Pauli twirling or zero-noise extrapolation \cite{Li2017, Temme2017} remain important for most experiments, these methods have been shown to be of limited use to extract noiseless expectation values \cite{Quek2024}. Significant research effort has gone into optimizing single- and multiqubit gate fidelities \cite{Li2023, Yan2018, Werninghaus2021, Hyypp2024}, however it is becoming increasingly clear that readout errors also pose an important obstacle to general-purpose quantum computing \cite{Mrton2023}. On the analog quantum simulation side, precise readout is just as important to prevent biasing of reconstructed observables. Therefore, recent research has focused on understanding \cite{Wilen2021, Tuziemski2023} and improving the multiplexed dispersive readout for superconducting qubits \cite{Walter2017, Heinsoo2018}. In this work we aim to contribute to understanding and mitigating readout errors with the help of quantum detector tomography \cite{Lundeen2008, Chen2019, Maciejewski2020, Maciejewski2021}. This approach assumes that state preparation in the quantum system is accurate, and it seeks to mitigate readout errors by reconstructing the experimental measurement operator instead of  assuming that the readout is accurately modeled by projective computational basis measurements. 

In particular, correlated readout errors have recently been observed in superconducting qubit systems. 
Refs.~\cite{Kono2024,Li2025} consider correlated bitflip errors and energy relaxation respectively. Both discuss physical mechanisms that may occur during readout, effectively manifesting as correlated readout, which would be captured by readout correlation coefficients. A further example could be fluctuations in the mutual electric ground. Numerical simulations of correlated readout errors from experiments and noise models have been reported in Ref.~\cite{Aasen2025}. Furthermore, Ref. \cite{deGroot2010} discusses methods to decrease electromagnetic crosstalk enabled readout correlations in superconducting flux qubits. Ref.~\cite{Heinsoo2018} provides experimental evidence for qubit dephasing due to correlated readout errors.

We apply a comprehensive readout-error-mitigation protocol, introduced in Ref.~\cite{Aasen2024}, to a multiqubit quantum state tomography experiment. The protocol mitigates both coherent and correlated readout errors, going beyond traditional techniques that address only classical readout errors \cite{Maciejewski2020, Geller2021, Geller2021-b} and enables very precise state reconstruction.  Our experiments are implemented on a superconducting qubit chip with four transmons, which we manipulate and read out using frequency-division multiplexing \cite{Jerger2012}. This architectural choice simplifies wiring, but it can be prone to significant readout crosstalk, and it is hence an ideal testbed for multiqubit readout-error mitigation. By employing a traveling-wave parametric amplifier (TWPA) \cite{Macklin2015}, we can perform single-shot readout of the qubits.
We introduce a readout quality metric that uses the number of shots necessary to reach a specified state reconstruction infidelity, quantifying the efficiency of extracting information from the system. The proposed metric takes into account coherent errors, also referred to as quantum errors, which we compare with assignment fidelities, defined as the probabilities of correctly assigning labels 0 or 1 to the prepared computational basis states.

Given the substantial number of repetitive measurements required for detector tomography, it is important to avoid spending time on overly precise calibration of the reconstructed positive operator-valued measures (POVMs).
 Therefore, we also investigate different distributions of a fixed shot budget between quantum detector tomography (QDT) and quantum state tomography (QST) to determine which results in the lowest achievable infidelity.
Furthermore, we demonstrate two- and three-qubit density matrix reconstruction that takes into account coherent and correlated readout noise \cite{Aasen2024}, which opens the way to reducing readout errors in complete benchmarking of two- and three-qubit gates. This is motivated by renewed interest in multiqubit entangling gates in the community \cite{Kim2022, Itoko2024, Li2024}.

An advantage of detector tomography is that it explicitly reconstructs the POVM corresponding to the multiqubit readout process. This enables extraction of multiqubit readout correlation coefficients of the system, regardless of their origin, which could include off-resonant drive of the resonator, mechanical vibrations \cite{Kono2024}, or ionizing radiation \cite{Thorbeck2023}. In multiplexed readout, a perfect measurement of a qubit should not depend on the state of its neighbors, which makes readout correlation a suitable indicator of readout quality \cite{deGroot2010}. 

The readout correlation coefficients should be independent of the state distinguishability in the readout plane. By lowering the readout pulse amplitude, effectively lowering state distinguishability, we experimentally demonstrate that readout correlation coefficients satisfy this property.

\section{Preliminaries}
\subsection{Generalized measurements}
To characterize a noisy detector, it is essential to adopt a framework that goes beyond projective measurement. This approach can be formulated by a set of operators possessing the following properties,
\begin{equation}
    M_i^\dag = M_i, \quad  M_i\geq 0, \quad\text{and} \quad  \sum_i M_i = \mathbb{1},
    \label{eq:POVM_definition}
\end{equation}
where $\boldsymbol{M}=\{M_i\}$ is the POVM with elements $M_i$. When taking the expectation values of the measurement operators, one can retrieve the probabilities of obtaining different outcomes $i$ in a measurement process, 
\begin{equation}
    p_i = \Tr(\rho M_i).
\end{equation}

\subsection{Detector tomography}
\label{sec:detector tomography}
Characterization of the noisy measurement device is achieved by performing detector tomography \cite{Lundeen2008}.  The goal of detector tomography is to reconstruct the generalized measurement operators $\boldsymbol{M}=\{M_i\}$ that satisfy the constraints in eq.~\eqref{eq:POVM_definition} through experimentally measured outcome probabilities 
\begin{equation}
    \frac{n_{is}}{N} = p_{is} = \Tr(\rho_s M_i),
    \label{eq:QDT_LI}
\end{equation}
where $N$ is the total number of measurements and $n_{is}$ is the number of outcomes $i$ observed from calibration state $\rho_s$.
To enable full detector tomography, the set of calibration states must be informationally complete. In our experiment we chose the symmetric and informationally complete tetrahedron states
\begin{equation}
    \begin{split}
        \ket{\psi_1} &= \ket{0},\\
    \ket{\psi_2} &= \frac{1}{\sqrt{3}} \ket{0} + \sqrt{\frac{2}{3}} \ket{1},\\
    \ket{\psi_3} &= \frac{1}{\sqrt{3}} \ket{0} + \sqrt{\frac{2}{3}} e^{i \tfrac{2\pi}{3}}\ket{1},\\
    \ket{\psi_4} &= \frac{1}{\sqrt{3}} \ket{0} + \sqrt{\frac{2}{3}} e^{i \tfrac{4\pi}{3}}\ket{1}.\\
    \label{eq:Tetrahedron_states}
\end{split}    
\end{equation}
With the full set of measurement probabilities in eq.~\eqref{eq:QDT_LI} one can estimate the most likely set of operators through a maximum-likelihood estimator, as outlined, for example, in~\cite{Fiurek2001}.

\subsection{Readout-error-mitigated quantum state tomography}
With a reconstruction of a noisy, experimentally measured POVM, it is possible to use the POVM to construct a readout-error-mitigated density matrix. The approach we use is described in Ref.~\cite{Aasen2024} and is based on directly integrating the full set of estimated measurement operators into a density matrix estimator. The protocol is specifically based on estimating the density matrix using the adapted likelihood function, 
\begin{equation}
\mathcal{L}_{\boldsymbol{M}^{\text{estm}}}(\rho)  \propto \Pi_i  \Tr(\rho M_i^{\text{estm}})^{n_i}=\left(\Pi_i (p_i)^{\hat{p}_i}\right)^N,
    \label{eq:Adapted_likelihood_function}
\end{equation}
with either a maximum-likelihood estimator \cite{Lvovsky2004} or a Bayesian mean estimator \cite{Struchalin2016}.

This approach allows for beyond-classical readout-error mitigation as it naturally accommodates coherent errors, and it does not require any channel inversion or numerical optimization while still guaranteeing physical density matrix reconstruction.

\subsection{Multiplexed qubit readout}
Multiplexed readout \cite{Jerger2012} is a central architectural choice for superconducting quantum processors. By coupling multiple resonators to a single transmission line, one can save circuit footprint on the chip. In conventional transmon processors, transmission lines account for a significant chip surface area. As resonators now share their transmission line, their readout pulses are sent through the cryostat on the same microwave line, and they are also amplified and mixed simultaneously after reading out the qubit state. This architectural choice needs fewer amplifiers and other microwave equipment. However, it has a significant drawback: resonators on a single transmission line cannot be too close in frequency, otherwise they will be measured with crosstalk. A readout pulse synthesized for one resonator will also drive a neighboring one, leading to a bias in the derived qubit state. Recently, deep-learning methods have been explored to increase single-shot assignment fidelity in multiplexed readout operations \cite{Lienhard2021}.

\subsection{Quantifying correlated readout errors}
A major challenge in scaling up quantum hardware is induced correlations and crosstalk, e.g. due to  multiplexed qubit readout. A useful tool for characterizing readout correlations was introduced in Refs.~\cite{Maciejewski2021, Tuziemski2023}, where, by using reconstructed two-qubit POVMs, asymmetric correlation coefficients $c_{j\rightarrow i}$ could be extracted. These coefficients quantify how much the measurement results from qubit $i$ are affected by the state of qubit $j$.
For a more detailed discussion of the definition of the correlation coefficients, see App.~\ref{app:correlation_coefficients}.

Since the correlation coefficients are asymmetric, we also define a symmetric version $c_{i\leftrightarrow j} = (c_{j\rightarrow i} + c_{i\rightarrow j})/2$. The coefficients can be modified to measure only the classical correlations, represented as $c^C_{i\leftrightarrow j}$, which are determined exclusively by the diagonal elements in the measurement operators. The full correlation coefficients contain the effect of both diagonal and off-diagonal elements, which we call quantum correlation coefficients, represented by $c^Q_{i\leftrightarrow j}$.

\subsection{Infidelity}
To quantify the quality of a reconstructed density matrix, we use the quantum infidelity, $I(\rho,\sigma)$, as a figure of merit. In the case of pure target states $\sigma$, the infidelity with respect to the estimated state $\rho$ is given by
\begin{equation}
    I(\rho, \sigma) = 1 - \Tr(\rho\sigma).
\end{equation}


\section{Experimental realization}
\subsection{Experimental setup}
\label{sec:experimental_setup}
\begin{figure}[t]
    \includegraphics[width=0.98\linewidth]{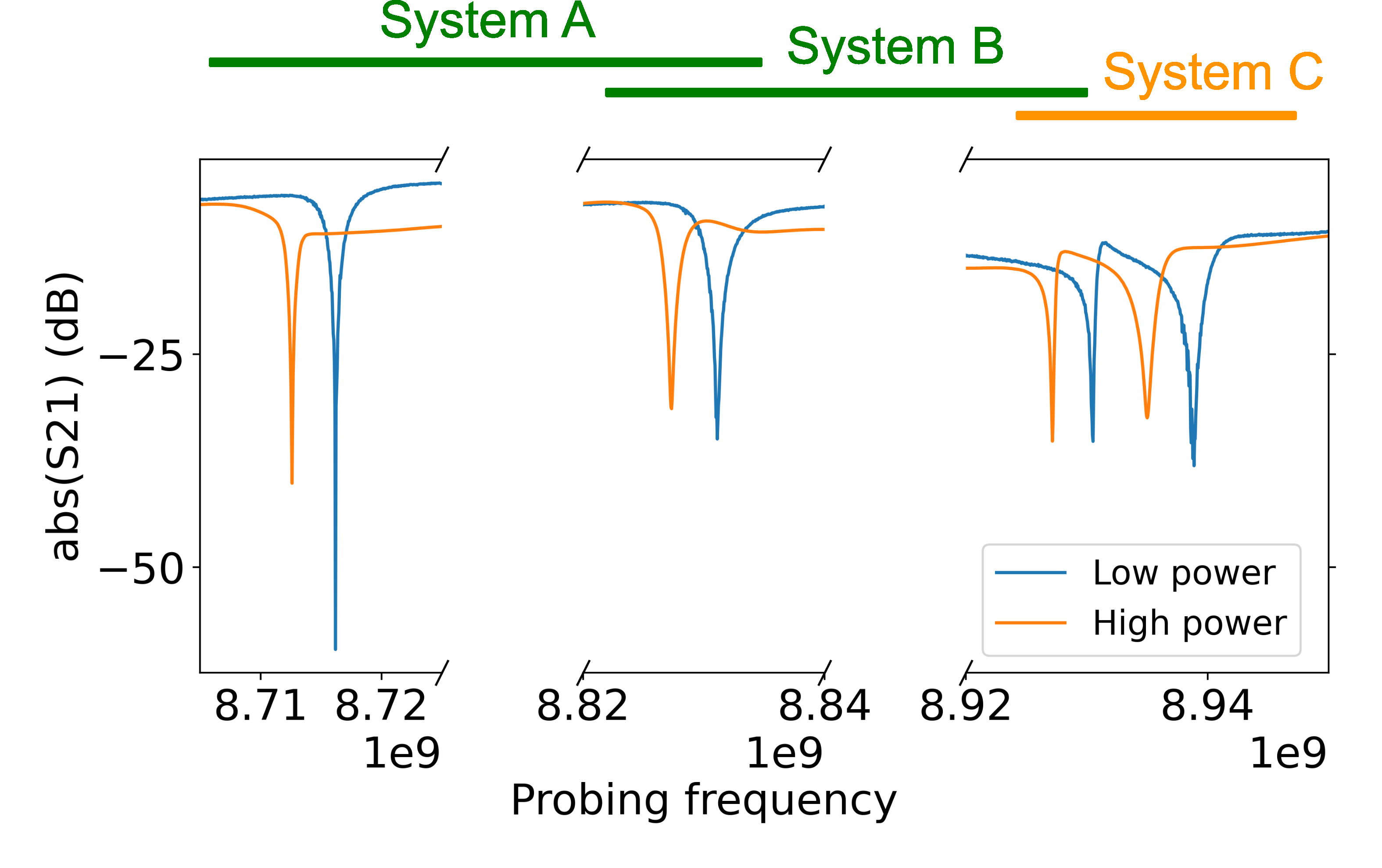}
    \caption{\justifying Resonator spectroscopy of the four-qubit system. At sufficiently low readout powers, each qubit couples coherently to its resonator, enabling dispersive readout. Systems A, B and C are two-qubit systems of neighboring resonator-qubit systems. An important feature is the closeness of resonators in System C, symbolized by the orange color. From left to right, the corresponding qubits are q0, q1, q2, and q3.}
    \label{fig:4q-spectroscopy}
\end{figure}

We investigate a device with four transmon qubits \cite{Barends2013}, each coupled individually to a resonator for readout. The device is measured below $15\, \text{mK}$ in a dry dilution cryostat. The frequencies of the qubits and resonators, as well as typical coherence times can be found in App. \ref{App: qubit parameters}. We define System A as the two-qubit subsystem comprising q0 and q1, System B as the subsystem comprising q1 and q2, and finally System C as the subsystem comprising q2 and q3, as shown in Fig. \ref{fig:4q-spectroscopy}. A description of the readout chain in the context of readout errors is found in App.~\ref{App: readout chain}.

\begin{figure}[t]
    \includegraphics[width=0.95\linewidth]{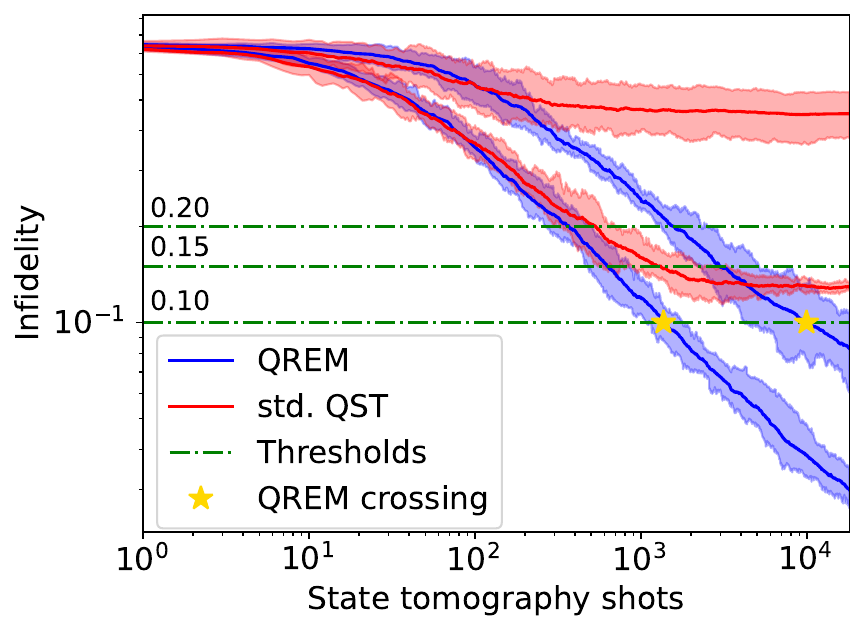}
    \caption{\justifying Mean infidelity curves crossing different threshold values (corresponding to infidelity values of 0.10, 0.15, and 0.20). Two different readout parameters are examined, where blue curves are with and red curves are without quantum readout-error mitigation (QREM). The curves are created using a Bayesian mean estimator of the reconstructed quantum states. Only readout-error-mitigated infidelities reliably cross the given thresholds, where the lowest threshold was only crossed by the QREM curves. Each curve is averaged over 16 Haar random initial states, and the shaded area indicates the interquartile range. }
    \label{fig:def infid}
\end{figure}

\subsection{Experimental protocol}
To reconstruct the POVM that describes the measurement process, we perform detector tomography. We prepare calibration states on the qubits using single-qubit gates. The calibration states are explicitly given in Sec. \ref{sec:detector tomography}. For the phase offset between the state preparation and readout control pulses, we use virtual $Z$ gates. The calibration states are measured in a set of different bases corresponding to the projectors onto the eigenbasis of the three Pauli operators $X$, $Y$ and $Z$. The POVM is then reconstructed in postprocessing according to Ref.~\cite{Aasen2024}.

Once the detector tomography has been completed, we prepare different quantum states to which we can apply quantum readout-error mitigation. The prepared states are randomly selected from the set of product states. Each single-qubit state is generated by uniformly sampling from the surface of the corresponding Bloch sphere, referred to as Haar random states. Haar random states are created by applying a random unitary matrix $U$, drawn at random from the unitary group $U(2^{n_\text{qubits}})$, to the ground state $\ket{0}$ \cite{mezzadri2006}.  

\subsection{Inducing readout noise}

To study the effect of readout noise on detector tomography and its expectation values, we induce noise that does not affect state preparation and takes effect only after the quantum state of interest has been prepared. The dominant noise sources we investigate stem from suboptimal (i.e. weaker or stronger than ideal) readout power of the resonators and insufficient amplification in the readout chain.

The amplitude of the readout pulses can be adjusted individually or simultaneously for multiple qubits. Generally, if the readout amplitude is too low, the measurement may not be fully projective, and reliable state distinguishability is not possible. In the case of very high readout amplitude, we excite states outside the qubit manifold, biasing the data \cite{Walter2017}. 

We sweep the frequency and amplitude of the continuous pump tone of the parametric amplifier. This results in varying amplification and therefore modifies the overlap of basis states in the IQ (in phase and quadrature signal) plane.

\begin{figure}[t]
    \includegraphics[width=.95\linewidth]{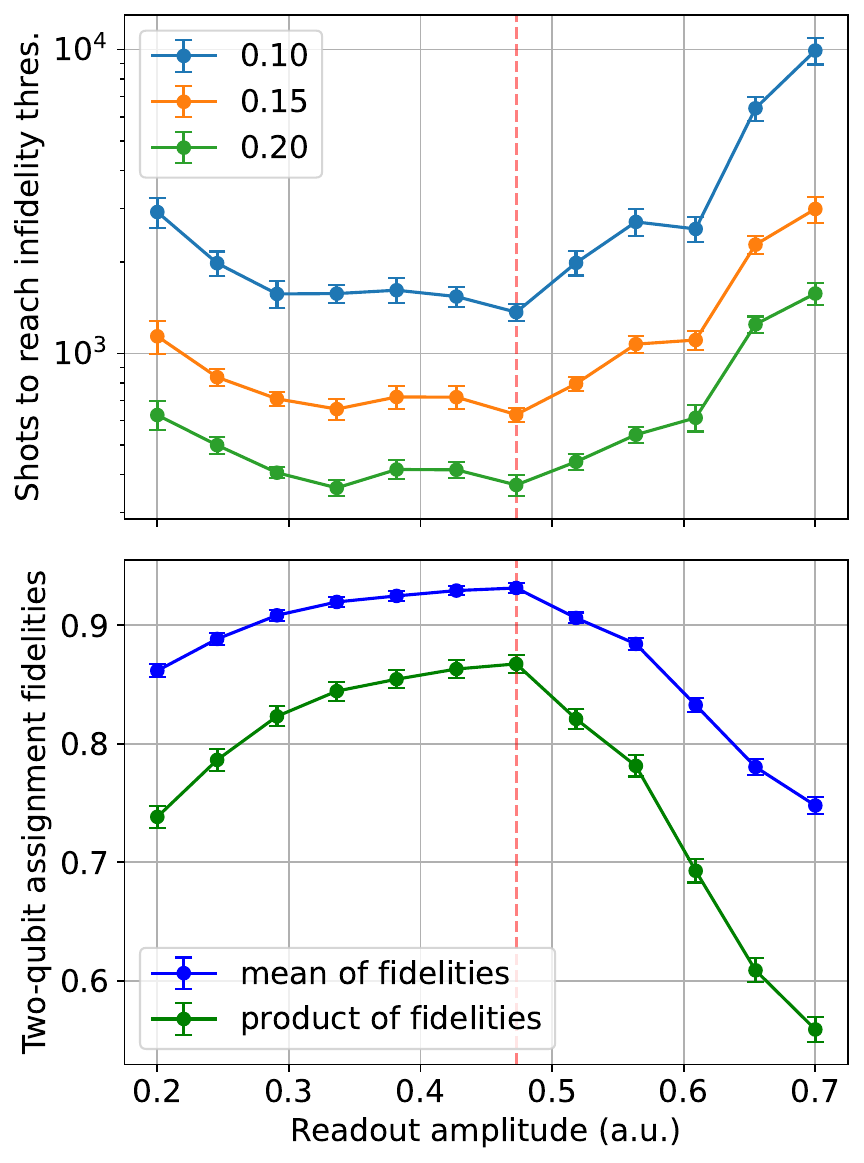}
    \caption{\justifying Comparison of quantum-state-tomography-based readout benchmark with assignment fidelities for varying readout amplitudes. The vertical dashed red line indicates optimal readout amplitude. \textbf{Top:} The number of two-qubit state tomography shots required to reach infidelity thresholds of 0.10, 0.15 and 0.20 for various readout amplitudes. The error bars are bootstrapped standard deviations. For very low readout powers, the distinguishability of the basis states is significantly reduced. For very high readout amplitudes, the measurement process excites higher states in the system, which the state classifier cannot correctly take into account. Each point uses 558 000 shots. \textbf{Bottom:} Mean and product of single-shot assignment fidelities. Each point uses 2000 shots.}
    \label{fig:2q-conv-vs-ropower}
\end{figure}

\section{Readout-error-mitigated multiqubit state tomography}
We apply multiqubit readout-error mitigation to the system described in the previous section. We use a generalization of the experimental protocol depicted in Fig.~1 of Ref.~\cite{Aasen2024} to multiqubit systems.
We study the rate of information extraction from readout-error-mitigated quantum state tomography experiments. The rate of information extraction serves a proxy for the total readout noise in the system, which includes both correlated and coherent errors.  We use this quantity as a quality metric for the readout, which we compare to assignment fidelity. Assignment fidelity is the standard readout quality metric, but it captures only readout errors in the computational basis.
In particular, we measure the number of experimental shots required to reach a given reconstruction infidelity threshold in Sec. \ref{sec:iv-A}. We then discuss the optimal distribution of a fixed shot budget between the different stages of the scheme in Sec.~\ref{subsec:shot_budget}. Finally, we apply the protocol to full three-qubit state tomography to assess the scalability of the approach (Sec.~\ref{sec:iv-B}).

\subsection{Extracting infidelity threshold crossings from Bayesian mean estimated state tomography}
We quantify the rate of information extraction by how many shots are required to reach a certain infidelity threshold in a state reconstruction experiment. To motivate the use of readout-error mitigation instead of unmitigated quantum state tomography, we investigate the infidelity curves of both cases in our two-qubit experiment. Fig.~\ref{fig:def infid} shows four mean infidelity curves, two quantum readout-error mitigated (QREM, blue), and two unmitigated (std. QST, red), as functions of the number of shots used for state tomography. Each curve is averaged over 16 products of single-qubit Haar random states \cite{yczkowski2011}. The crossing points of these curves with a given constant infidelity ($0.20, 0.15$, or $0.10$) are extracted using a Bayesian mean estimator \cite{Aasen2024}. This allows one to identify the exact number of shots at which the infidelity curve crosses a specified infidelity threshold value. We note that the unmitigated curves saturate at high infidelity values. In fact, at high levels of noise, they do not even cross these thresholds (green horizontal lines). This makes them inappropriate for precisely benchmarking readout quality. If one uses the readout-error-mitigated infidelity curves (depicted in blue) instead, the infidelity thresholds are crossed in the steady exponential regime (specifically, in the linear segment of the log-log plot) of the curves, making them suitable for benchmarking. Furthermore, infidelity curves are most accurately obtained when employing a Bayesian mean estimator, highlighting an advantage over maximum-likelihood estimators in the reconstruction process. Bayesian mean estimators can be efficiently computed for two qubits, after which, maximum-likelihood estimators gain an additional advantage in speed.

\subsection{Readout noise affecting information-extraction rate}
\label{sec:iv-A}
To investigate how the rate of quantum information extraction is influenced by readout noise, we perform readout-error-mitigated two-qubit quantum state tomography experiments on System A (see Fig.~\ref{fig:4q-spectroscopy}) with induced readout noise. We compare the results with a corresponding analysis using assignment fidelity as a readout quality metric. Since assignment fidelity depends exclusively on computational basis states and does not fully benchmark readout fidelity, our analysis aims to confirm the suitability of using assignment fidelity as a measurement standard.

\subsubsection{Varying readout power}
\label{sec:iv-B}
 We vary the readout power for both resonators simultaneously, so that they are equal at all times. This leads to the possible existence of multiple optima for information extraction.
The upper figure in Fig.~\ref{fig:2q-conv-vs-ropower} illustrates the number of shots necessary to achieve the infidelity thresholds.
The plot shows one optimum for information-extraction rates at $0.48 \,\text{(a.u.)}$ readout amplitude, illustrated by a red dashed line.
This is in agreement with both the product and the sum of single-qubit assignment fidelities (see the bottom panel in Fig.~\ref{fig:2q-conv-vs-ropower}). The optimum obtained for infidelity thresholding is statistically significant, whereas the measured assignment fidelities are also consistent with a plateau between readout amplitudes of 0.35 and 0.48. This is due to the relatively large error bars, limited by the number of computational basis states measured. It is expected that repeating the experiments with more shots would also highlight one clear optimum at 0.48 (a.u.) readout power.
The general features of the readout parameter landscape are explained by low state distinguishability at low readout powers and state leakage at higher readout power, as also observed, for example, in Ref.~\cite{Walter2017}. Both the assignment fidelity and infidelity convergence landscapes are similar and well explained by the above reasoning. The fraction of leaked states at higher readout power is plotted in App. \ref{app:leakage}.
In summary, the experiment demonstrates that the more comprehensive infidelity thresholding aligns with assignment fidelities at the optimal readout power for two-qubit multiplexed readout. Given the higher cost associated with infidelity thresholding without a corresponding increase in accuracy, assignment fidelity is preferred in this setup.

\subsubsection{Parametric amplification}
\begin{figure}[t]
    \includegraphics[width=0.95\linewidth]{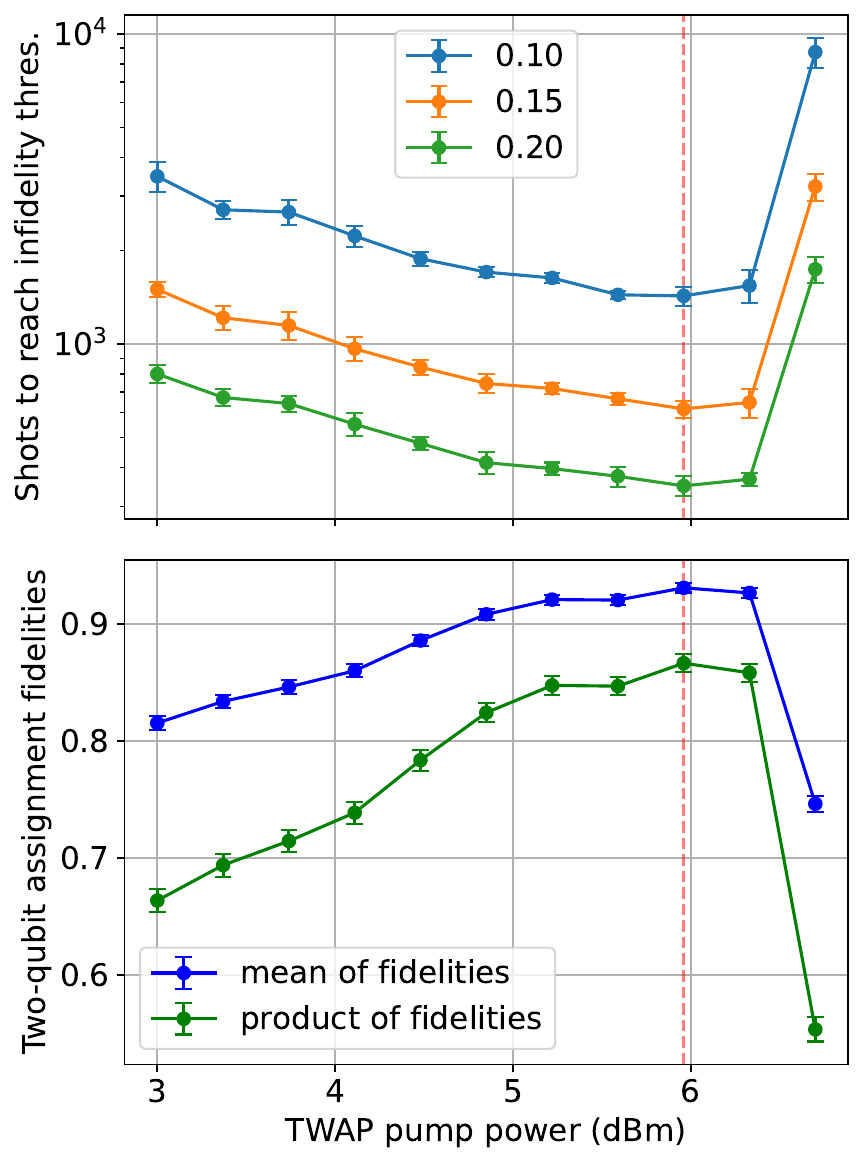}

    \caption{\justifying Comparison of quantum-state-tomography-based readout benchmark with assignment fidelities for varying amplification. The vertical dashed red line indicates optimal pump power. \textbf{Top:} The number of shots in state tomography needed to reach 0.10, 0.15, and 0.20 reconstruction infidelity. The error bars show bootstrapped standard deviations. Each point uses 558 000 shots. \textbf{Bottom:} Sum and product of assignment fidelities for the performed experiments. State separation increases with pump power up to approximately $6.5\,\text{dBm}$, after which it drops sharply. The error bars show propagated standard deviations from the assignment fidelity. Each point uses 2000 shots.}
    \label{fig:inf threshold vs twpa power}
\end{figure}

The resonators of System A are detuned by about $115\, \text{MHz}$ from each other, therefore it is expected that they could have different optimal operating parameters on the shared TWPA. This is due to the nontrivial frequency dependence of the gain, as measured in Ref.~\cite{Macklin2015}. Thus, it is not sufficient to test the TWPA on a single qubit. Performing state tomography on this system, see Fig.~\ref{fig:inf threshold vs twpa power}, we see that information extraction is fastest at around $6.0\,\text{dBm}$ pump power. This is in good accordance with the maximum of both the sum and product of single-shot assignment fidelities. Once the TWPA is oversaturated in pump power, a significant  slowing of information extraction by about a factor of 6 compared to the optimum is observed (for a more detailed measurement of this saturation, see App.~\ref{app:twpa_wider_range_fids}). At low pump powers, the information-extraction rate is slowed because of insufficient distinguishability of the computational basis states at readout, which can also be observed in the corresponding assignment fidelities.

Similar to varying the readout power, both infidelity thresholding and assignment fidelity agree on the optimal pump power.


\subsection{Optimal shot budget distribution}
\label{subsec:shot_budget}
Experiments often have to be limited in run time. This can be due to experimental drift or simply resource sharing. This raises the question of how one should optimally allocate a fixed measurement budget between detector tomography and state tomography to minimize the resulting infidelity in the reconstruction of a quantum state.
We define the shot ratio as
\begin{equation}
   r = \frac{\text{QDT shots}}{\text{QDT shots + QST shots}},
   \label{eq:ratio definition}
\end{equation}
i.e. the ratio of the shots used for detector tomography to the total number of shots performed in the reconstruction of a single state.

To find the optimal ratio for our experiment, we performed single- and two-qubit tomography with a fixed number of total shots for different values of $r$. For a representative convergence of the state reconstruction, an average of the same 40 Haar random states was used for each value of $r$ for single-qubit reconstruction. For two-qubit shot budgeting, ten products of single-qubit Haar random states were used.

\begin{figure}[t]
\includegraphics[width=1\linewidth]{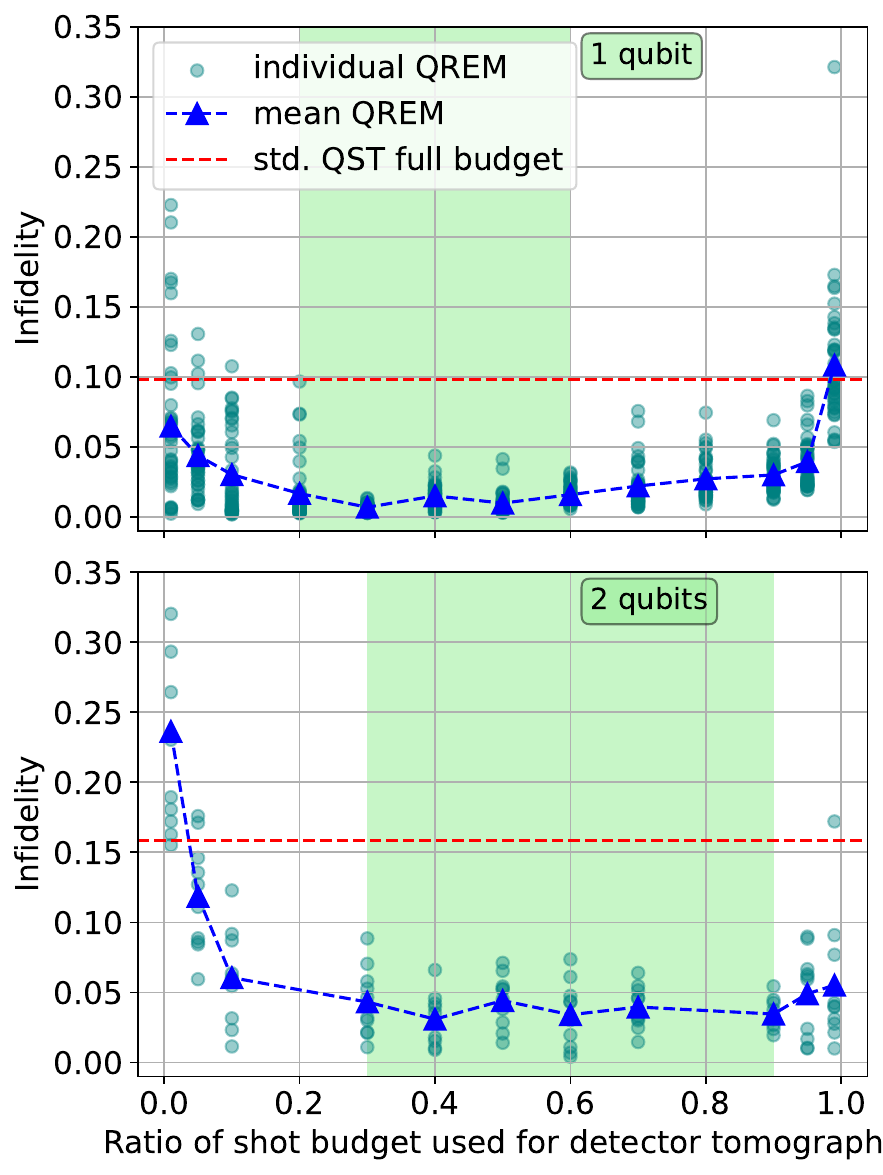}
    \caption{\justifying Reachable infidelities for different distributions of a given shot budget in readout-error-mitigated state reconstruction. The plotted infidelities refer to the final infidelity at the end of quantum state reconstruction. The blue triangles correspond to  the mean infidelities, while the spread in each case is shown with teal circles. The red horizontal line indicates the mean final infidelity at which the full budget was used for  standard state tomography. The identified optimal shot ratio is marked in green. \textbf{Top:} Single qubit, shot budget of 10 000, averaged over 40 Haar random states. \textbf{Bottom:} Two-qubit reconstruction with a shot budget of 40 000, averaged over ten Haar random states. }
    \label{fig:shot budgets}
\end{figure}

The results of the experiments can be seen in Fig.~\ref{fig:shot budgets}. For both single- and two-qubit readout-error-mitigated state tomography, spending half the shot budget on calibration (QDT) is a reasonable choice. In the case of very few shots being spent on calibration, the resulting infidelity is limited by sample fluctuations in the estimated measurement operator. On the upper end, if too few shots remain for state tomography, the experiment will not reach the lowest infidelities. The ideal shot ratio ranges are highlighted in green in Fig.~\ref{fig:shot budgets}. From these plots, we can estimate the optimal value of $r$, as defined in eq. \eqref{eq:ratio definition}, to be $$r_1 = 0.4 \pm 0.2$$ for a single qubit and $$r_2 = 0.6 \pm 0.3$$ for two-qubit experiments. In App.~\ref{app:theory_shot_budgeting}, we consider a simple theoretical model based on the number of independent parameters to be estimated. Our predicted optima are at larger values than the observed ones, indicating that there is a significant amount of structure which likely simplifies the statistical estimation task. 


\subsection{Three-qubit state tomography}
\begin{figure}[t]
    \includegraphics[width=0.98\linewidth]{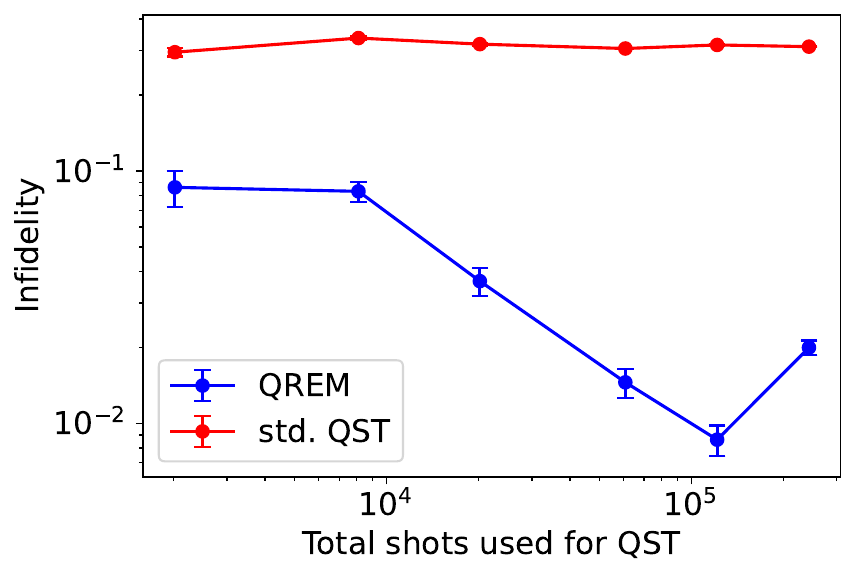}
    \caption{\justifying Reconstruction of a three-qubit density matrix with and without quantum readout-error mitigation (QREM) as a function of the total number of shots used for QST. The error bars indicate the standard deviation of the estimator obtained from 1000 bootstrap resamplings. }
    \label{fig:3q infidelity vs shots}
\end{figure}

In light of the renewed interest in three-qubit gates, noted for their potentially higher fidelity, shorter duration, and ability to replace multiple two-qubit gates \cite{Kim2022, Warren2023, Itoko2024},  we perform
three-qubit quantum state tomography  with and without readout-error mitigation. This examination aims to evaluate the scalability of the protocol introduced in Ref.~\cite{Aasen2024} to more qubits. The experiment was repeated on the same prepared state with an increasing number of shots per basis measurement in both detector tomography and state tomography. The results are plotted in Fig.~\ref{fig:3q infidelity vs shots}. Infidelities below $0.01$ were reached, with an outlier data point at 12000 shots per basis measurement likely attributable to measurement drift, which we investigate further in App.~\ref{sec:drift}. A comparison of the theoretically expected density matrix to the reconstructed one with and without readout-error mitigation can be found in App.~\ref{App: 3q qst} Fig.~\ref{app:3q-density-matrix}.
We observe an infidelity improvement of around a factor of $30$ when comparing state tomography with and without readout-error mitigation. Therefore, we conclude that the method introduced in Ref.~\cite{Aasen2024} is scalable to three superconducting qubits and is capable of significantly reducing readout errors. This provides a way to benchmark and understand three-qubit gates, even in the presence of coherent readout errors in the system. Complete state reconstruction is intrinsically not a scalable method. However, with the established applicability to three-qubit benchmarking, the technique can be applied to characterize few-qubit subsystems within larger qubit systems, which is typically sufficient for mitigation of correlated errors \cite{Aasen2025, Geller2021-b, Tuziemski2023}.

\section{Two-qubit correlations from detector tomography}
As multiplexed readout has become widespread in superconducting qubit devices of growing size, it  is important to reliably quantify potential readout correlations that exist due to limited resonance frequency spacing. 
In this section we use correlation coefficients extracted from detector tomography to investigate our qubit system. In addition, we analyze the dependence of these correlation coefficients on the amplitude of the readout pulse.

\subsection{Correlation measurements for two-qubit subsystems}
\begin{figure}[t]
    \includegraphics[width=\linewidth]{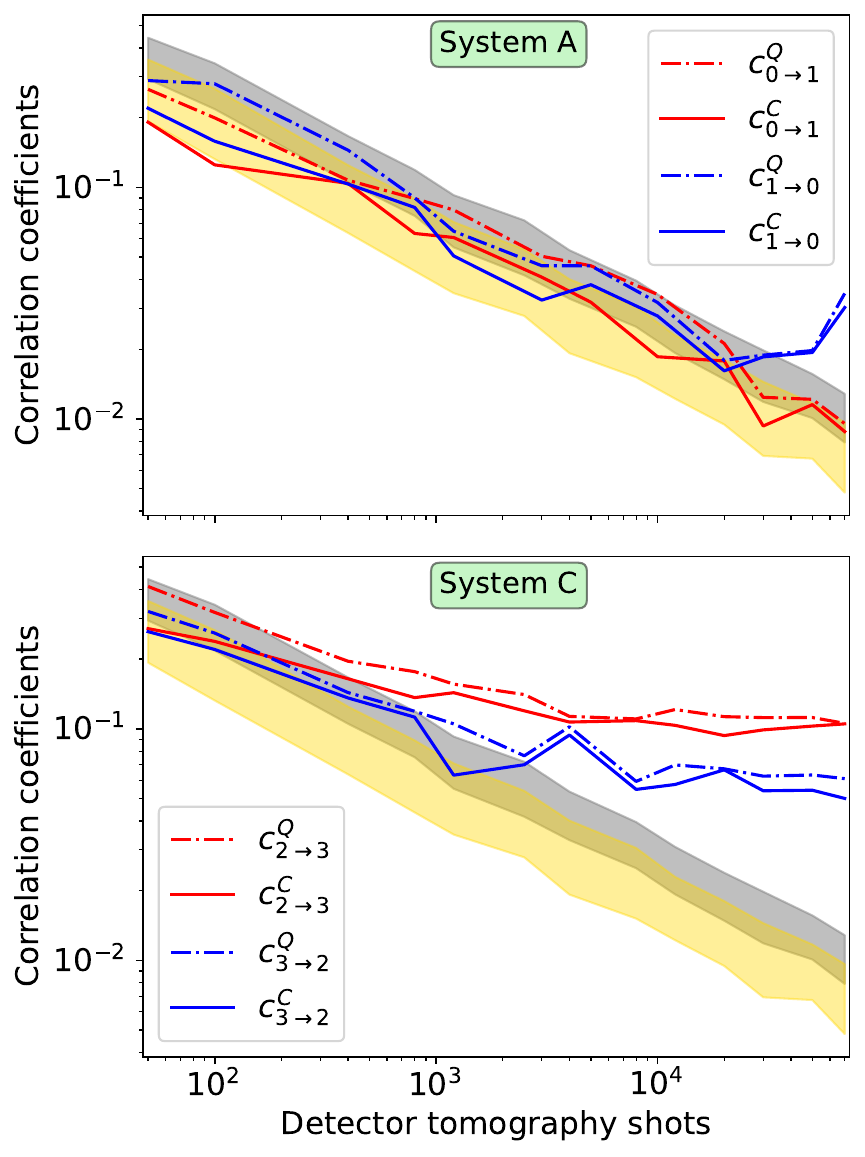}
    \caption{\justifying Correlation coefficients plotted as a function the total number of detector tomography shots used for reconstructing the measurement operators. The silver and gold shaded areas represent the standard deviations of quantum and classical correlation coefficients, respectively, derived from numerical simulations of measurements without correlation. Curves that fall within these regions indicate 
    compatibility with uncorrelated readout. \textbf{Top:} In System A, the readout correlations do not exhibit any readout correlations within the bounds of the experimental constraints. \textbf{Bottom:} System C displays nonzero asymmetric correlations, which suggests that readouts are indeed correlated within the system. However, no coherent correlations were observed within the limitations of the experiment.}
    \label{fig:q0q1 and q2q3 correlations vs dt shots}
\end{figure}

To investigate the effect of resonator frequency spacing, we performed detector tomography on Systems A and C (see Sec.~\ref{sec:experimental_setup}) and extracted both quantum and classical correlation coefficients as functions of detector tomography shots. The results are plotted in Fig.~\ref{fig:q0q1 and q2q3 correlations vs dt shots}. At low numbers of shots, the correlation coefficients are  compatible with uncorrelated POVMs (shaded silver and gold areas), indicating that the coefficients are dominated by sample fluctuations. System A is always compatible with uncorrelated readout, while System C shows strong asymmetric readout correlations. At high numbers of shots for System A, the $0\rightarrow1$ correlation coefficients move outside the standard deviation of uncorrelated POVMs, but due to low statistics, we consider this to be consistent with uncorrelated POVMs. These observation are consistent with the observed frequency spacing of resonators in the two systems: $\delta_{B} = 100 \, \text{MHz}$ vs $\delta_{C}= 8\, \text{MHz}$. Neither system exhibits significant correlated coherent readout errors. In System C, the correlation coefficients are of the order of 0.1, comparable to those reported for the Rigetti Aspen-M-3 in Ref.\cite{Tuziemski2023}. Ref.\cite{Aasen2025} presents numerical simulations of noisy readout using both experimental and synthetic systems. Fitting the probabilistic iSWAP channel $\mathcal{E}(M_i) = (1-k)M_i + k U_{\text{iSWAP}} M_i U_{\text{iSWAP}}^\dagger$ to the observed correlation coefficients in System C is consistent with a noise strength of approximately $k \approx 0.1$.

We therefore conclude that the correlation coefficients can be used as a precise tool to test whether resonators are sufficiently far from each other both spatially on the chip and in the frequency domain.
As correlations are dependent on in-situ readout variables such as readout power, it is possible to optimize these parameters to maximize the rate of information extraction while simultaneously minimizing the correlations between different qubit readouts. Our results therefore open the way for a more comprehensive optimization of multiqubit readout.

\subsection{Two-qubit correlation dependence on readout noise strength}
\label{sec:correlation_under_state_distinguishability_noise}
To better understand how the correlation coefficients behave under additional noise sources, we investigated how the coefficients change when lowering the distinguishability of computational basis state using readout pulses with varying amplitudes. Lowered state distinguishability can be modeled as depolarizing-type noise on each qubit, which is not expected to impact the correlation coefficients, see App.~\ref{app:corr_coeff} for more details.

We define the relative readout amplitude as the ratio between the used amplitude of the readout tone and its ideal amplitude. To avoid leakage to higher levels, we restrict ourselves to decreased readout amplitudes, corresponding to relative readout amplitudes not larger than $1$. Ad extremum, the single-shot assignment fidelity measured with a relative readout amplitude of $0.1$ is only about $55\%$ compared to around $95\%$ at a readout ratio of $1$.

\begin{figure}[t]
    \includegraphics[width=\linewidth]{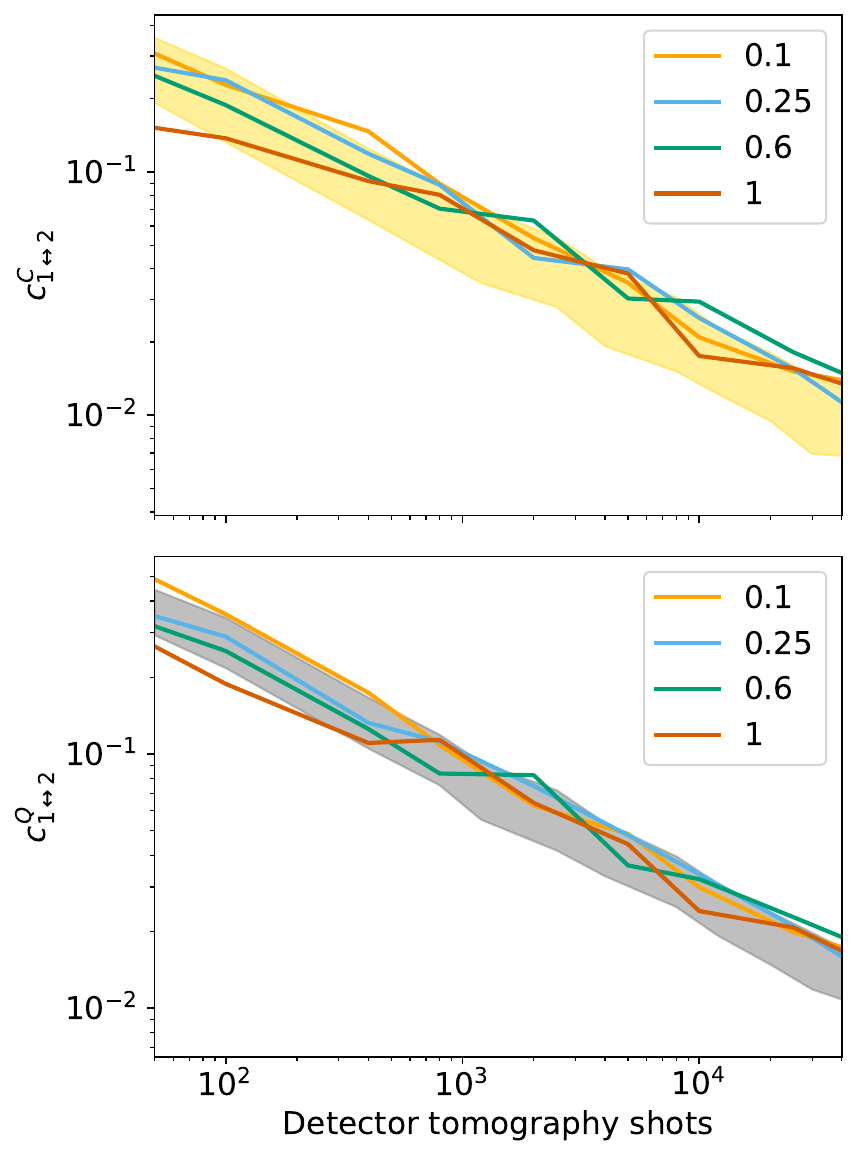}
    \caption{\justifying Correlation coefficients in System A as a function of the number of detector tomography shots for lowered relative readout amplitudes. The silver and gold shaded areas represent the standard deviations of quantum and classical correlation coefficients, respectively, derived from numerical simulations of measurements without correlation. \textbf{Top:} Classical correlations decrease independently of the relative readout amplitude, remaining consistent with uncorrelated readout. \textbf{Bottom:} Quantum correlations have larger values but are always consistent with uncorrelated readout.}
    \label{fig:correlation-vs-noise}
\end{figure}

The results of the experiments can be seen in Fig.~\ref{fig:correlation-vs-noise}. Within the bounds of the experiment, both quantum and classical correlations are independent of the strength of the readout noise, even at very high levels of noise, confirming our theoretical expectations. 

\section{Conclusions and outlook}
We investigated a generalized method for readout-error mitigation based on detector tomography and explored its utility as a benchmarking tool for superconducting qubits. The protocol was applied to various two- and three-qubit state reconstruction experiments with superconducting transmon qubits. We introduced a benchmark for qubit readout quality that goes beyond conventional techniques focused on the fidelity of computational basis state assignments, which primarily detect classical errors. Our approach uses quantum infidelities between prepared states and readout-error-mitigated reconstructed states as a proxy for the information-extraction rate. Using  reconstruction infidelity characterizes both coherent and correlated readout errors, making it a more comprehensive benchmark than assignment fidelity. We thus propose using the rate of infidelity threshold crossings as a validation tool for assignment fidelity and as a suitable comprehensive figure of merit for certifying single- and multiqubit readout processes. In particular, when optimizing the readout chain, assignment fidelity fails to properly capture correlated readout errors, while infidelity-thresholding-based benchmarking is able to quantify the correlations present in the measurement process. Therefore, if assignment fidelity falls short, this new figure of merit can be used to find optimal operating parameters for readout chains, such as fine-tuning amplifiers for ideal readout performance. We studied the impact of different readout parameters on the rate of information extraction. In particular, the dependence on the pump power of the parametric amplifier and the amplitude of the readout resonator drive tone were investigated. The assignment fidelity- and infidelity-convergence-based methods produced very similar results in the investigated parameter range, validating the use of assignment fidelity in the investigated system. 

We investigated how the budgeting of shots between detector tomography and state tomography influenced the state reconstruction infidelity and thus identified the optimal distribution of shots. To address the increasing interest in two- and three-qubit gates for superconducting qubits, we demonstrated that beyond-classical readout error-mitigation is suitable for these system sizes.

By performing readout correlation measurements, we showed that readout correlations are significant in the weakly detuned system. We observed no correlated coherent readout errors in this system. By adjusting the readout amplitude of the resonators, we showed that even high noise levels do not induce readout correlations, validating correlation coefficients as an additional useful benchmark for the multiplexed readout of superconducting qubits. 
Our comprehensive benchmark for quantifying qubit readout precision can be applied to test novel qubit architectures \cite{Kreuzer2025} and readout methods \cite{Gunyh2024}. The optimization of readout parameters should not only focus on improving the performance of individual qubits, but also on keeping readout correlations in multiplexed readout to a minimum. It would be interesting to perform optimization experiments on weakly detuned resonator systems to test this limit of qubit readout optimization.  Importantly, we encourage the convergence of reconstruction infidelity to be used in upcoming superconducting qubit experiments that aim to push the boundaries of readout accuracy. Furthermore, it would be interesting to compare the infidelity-thresholding approach to a similar approach that uses QREM that only mitigates correlated classical errors, as these approaches require fewer shots for calibration and could be a more economical solution if coherent errors can be neglected.

\section*{Data availability}
The experimental data are available upon request from A. Di Giovanni. The code and data used for the analysis presented are available in Ref.~\cite{Aasen2025Github}.

\acknowledgments
The authors thank Andy Ding for providing the qubit chip used in the study.
We also thank W. Oliver and G. Calusine for providing the parametric amplifier.
This work was partially financed by the Baden-Württemberg Stiftung gGmbH.

\section*{Author contributions}

A.D.G. performed the experiments with supervision from H.R., J.L, and A.U. The theory and numerics were implemented by A.S.A. with supervision from M.G. A.D.G. and A.S.A. performed the data analysis. The first manuscript was prepared by A.D.G. and A.S.A. All authors contributed to the finalization of the manuscript.

\section*{Competing interests}
The authors declare no competing interests.

\section*{Appendix}

\appendix

\section{Details on correlation coefficients}
\label{app:correlation_coefficients}
In Refs.~\cite{Maciejewski2021, Tuziemski2023} introduced a metric for quantifying how the state of one qubit impacts the state of another. This is based on efficiently measuring and reconstructing all possible two-qubit POVMs with detector overlapping tomography \cite{Maciejewski2021, Cotler2020}. A readout correlation measure can be obtained as follows. For each two-qubit POVM, trace-out half of the POVM conditioned on a state being measured in the traced-out subsystem. Optimize over all possible states present on the traced-out subsystems and select the largest distance between two traced-out POVMs. These correlation coefficients can be written explicitly as
\begin{equation}
    c_{j\rightarrow i} = \sup_{\rho_j, \sigma_j} ||(M^{i, \rho_j}_0 - M^{i, \sigma_j}_0)||_2,
    \label{eq:quantum_correlation_coefficients}
\end{equation}
where $M_{x_i}^{i, \sigma_j} = \sum_{x_j} \Tr_j(M^{ij}_{x_ix_j}(\mathbb{1}_i \otimes \sigma_j))$, $x_i$ and $ x_j$ label the outcomes on the two-qubit system, and $||A||_2$ refers to the spectral norm (i.e. the largest singular value of $A$). This corresponds to the correlation coefficients employed in Ref.~\cite{Tuziemski2023} using the ``worst-case" distance, with the difference that in that work, the norm applied is the infinity operator norm, $||A||_\infty =\text{max}_{1\leq k \leq d} \sum_{l=1}^d|a_{kl}|$, where $d=\dim(A)$. The coefficients quantify the impact of the state of qubit $j$ on the effective POVM measured on qubit $i$. Due to the asymmetry of these coefficients, it is useful to define the symmetric correlation coefficient $c_{i\leftrightarrow j} = (c_{j\rightarrow i} + c_{i\rightarrow j})/2$.

If the optimization in eq.~\eqref{eq:quantum_correlation_coefficients}  is restricted to only the computational basis states, i.e.~products of eigenstates of the $\sigma_z$ operator $\{\ket{0}, \ket{1}\}^{\otimes N}$, the coefficients are called classical correlation coefficients $c^C$. If the optimization is performed over all pure states \cite{Tuziemski2023}, the resulting coefficients are called quantum correlation coefficients $c^Q$. The quantum correlation coefficients cannot be smaller than the classical ones and are only equal if there are no coherent errors present in the readout process (up to statistical fluctuations).

\section{Qubit and resonator parameters}
\label{App: qubit parameters}

Qubit and resonator parameters varied slightly between different cooldowns. Some representative values are listed in Tables \ref{table:coherence-table1} and \ref{table:coherence-table2}.

\begin{table}[h!]
\centering
 \begin{tabular}{||c c c||} 
 \hline
 Qubit& $\omega_{res}$ ($\,GHz$) & $\omega_{01}$ ($\,GHz$)  \\  
 \hline\hline
 q0 & 8.716  & 4.455   \\ 
 q1 & 8.831  & 5.409   \\
 q2 & 8.931  & 4.127  \\
 q3 & 8.939  & 4.926  \\
 \hline
 \end{tabular}
  \caption{Qubit names, resonator- and qubit frequencies of the quantum device. The qubits are named in order from left to right on the physical chip, starting from q0. Qubit and resonator frequencies typically varied up to a few MHz between cooldowns. }
   \label{table:coherence-table1}
\end{table}

\begin{table}[h!]
\centering
 \begin{tabular}{||c c c c||} 
 \hline
 Qubit& Assignment fidelity (\%) & $T_1$ $(\mu s)$  & $T_2$ $(\mu s)$   \\  
 \hline\hline
 q0 & 96 & 118 & 67   \\ 
 q1 & 93 & 22 & 30 \\
 q2 & 92 & 27 & 56  \\
 q3 & 80  & 23 & ca. 15 \\
 \hline
 \end{tabular}
  \caption{Typical coherence times and single-shot readout fidelities of the individual qubits. The assignment fidelity values show the best possible single-qubit readout assignment fidelity. Multiqubit assignment fidelities were observed to be lower in some experiments.}
    \label{table:coherence-table2}
\end{table}

\section{Readout chain}
\label{App: readout chain}
To understand the origin of readout errors, we provide a brief review of the readout chain used for typical dispersive-shift measurements of superconducting qubits.
After reaching the transmission line on the chip, the readout signals couple into the resonators that match their frequency. The frequency of the resonator depends on the collapsed qubit state through the dispersive shift \cite{Zhu2013}. This signal is amplified first on the mixing chamber stage with a Josephson traveling-wave parametric amplifier (JTWPA), driven by a continuous microwave pump tone, which is also channeled into the remaining chain. Some signal amplitude is lost before and after the JTWPA due to nonsuperconducting cabling and connectors. The resulting signal enters two high-electron-mobility transistor (HEMT) amplifiers, which add noise photons to the signal. External mixers leak sidebands into the signal before it reaches the digitizer and integration module, finally resulting in a measured value of 0 or 1.

\section{Leakage and assignment fidelity}
\label{app:leakage}
In this section, we discuss the comparison of infidelity thresholding and assignment fidelity for varying readout powers in more detail (see Sec. \ref{sec:iv-B}).

\begin{figure*}[t]

    \includegraphics[width=.95\linewidth]{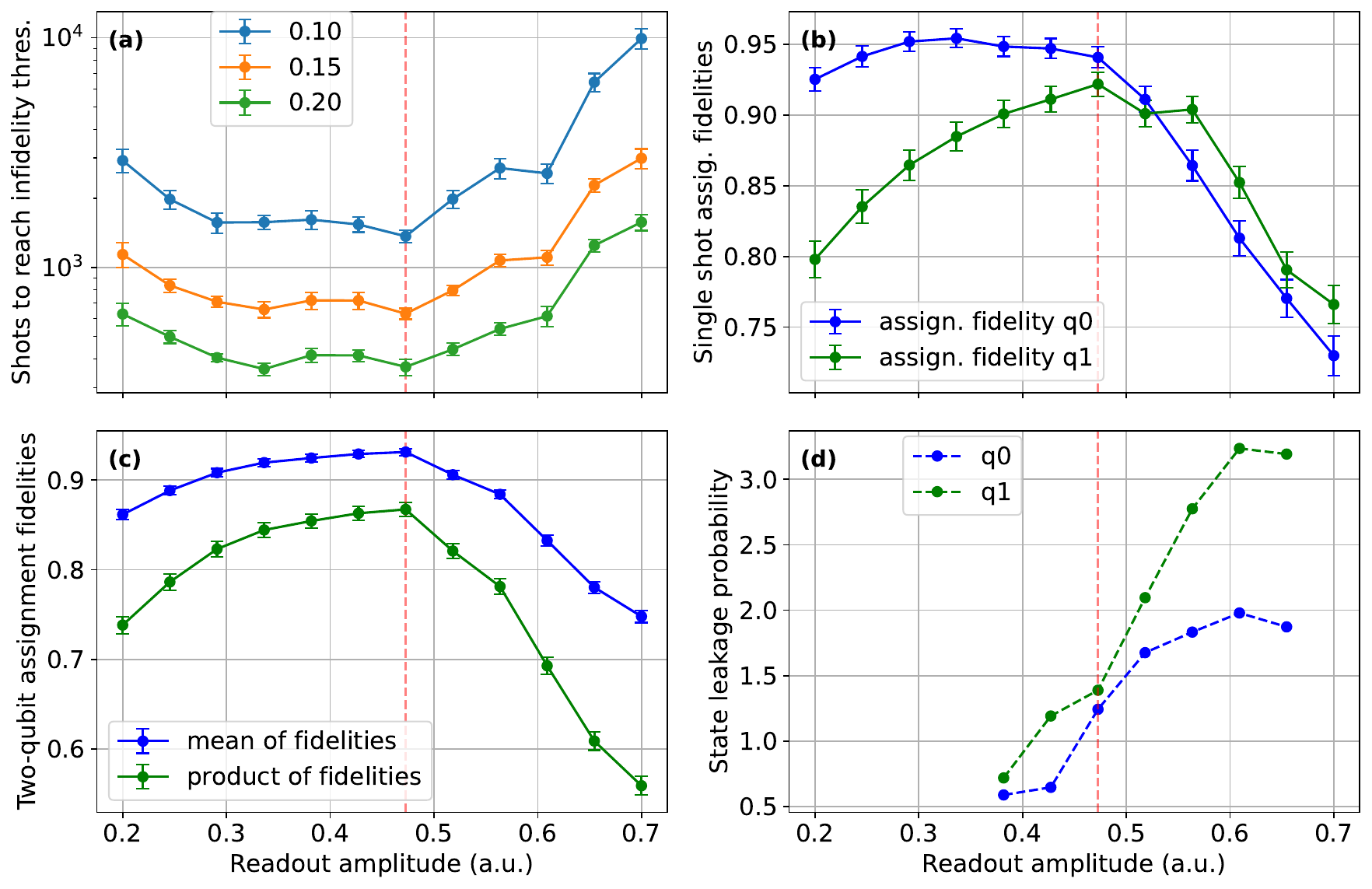}
    \caption{\justifying Comparison of readout-error-mitigated quantum-state-tomography-based readout benchmark with assignment fidelities. The vertical dashed red line indicates the optimal readout amplitude identified in Fig.~\ref{fig:2q-conv-vs-ropower}. \textbf{(a)} Number of two-qubit state tomography shots required to reach infidelity thresholds of 0.10, 0.15, and 0.20 for various readout amplitudes. The error bars are bootstrapped standard deviations. For very low readout powers, the distinguishability of the basis states is significantly lowered. For very high readout amplitudes, the measurement process excites higher states in the system, which the state classifier cannot correctly take into account. \textbf{(b)} Single-shot assignment fidelities during the experiments, each point uses 2000 shots. The error bars show one standard deviation. \textbf{(c)} Mean and product of single-shot assignment fidelities from Fig.~\ref{fig:2q-conv-vs-ropower}. 
    \textbf{(d)} Probability of measuring a leaked state as a function of readout amplitude.}
    \label{app:2q-conv-vs-ropower}
\end{figure*}

We treat the transmons as three-level quantum systems (qutrits) and distinguish three regions in the IQ plane, corresponding to the ground and excited states and the third (leaked) state. We prepare an equal number of $\ket{0}$ and $\ket{1}$ states and apply readout pulses with growing amplitudes to each qubit. The resulting distributions in the IQ plane are grouped into three clusters (ground, excited, and leaked), and a linear sum assignment is performed to relabel the clusters in maximum accordance with the known (but sometimes incorrect) labels 0 or 1. Note that the lowest readout power at which we can reliably perform the leakage estimate in this way is $1.5\times$ the optimal readout power. This is, because at lower readout powers, it is more difficult to reliably estimate the leakage ratio, as the dataset becomes very unbalanced, leading to misinterpretation (overfitting). This is a known problem for $N$-clustering algorithms.
The results are shown in Fig. \ref{app:2q-conv-vs-ropower}.
Although q1 is more prone to leakage, its assignment fidelity is less sensitive to this effect at high readout powers. This effect can be attributed to the position of states in the IQ (in phase-quadrature) plane.

\section{TWPA power saturation}
\label{app:twpa_wider_range_fids}
\begin{figure}[t]
    \includegraphics[width=0.95\linewidth]{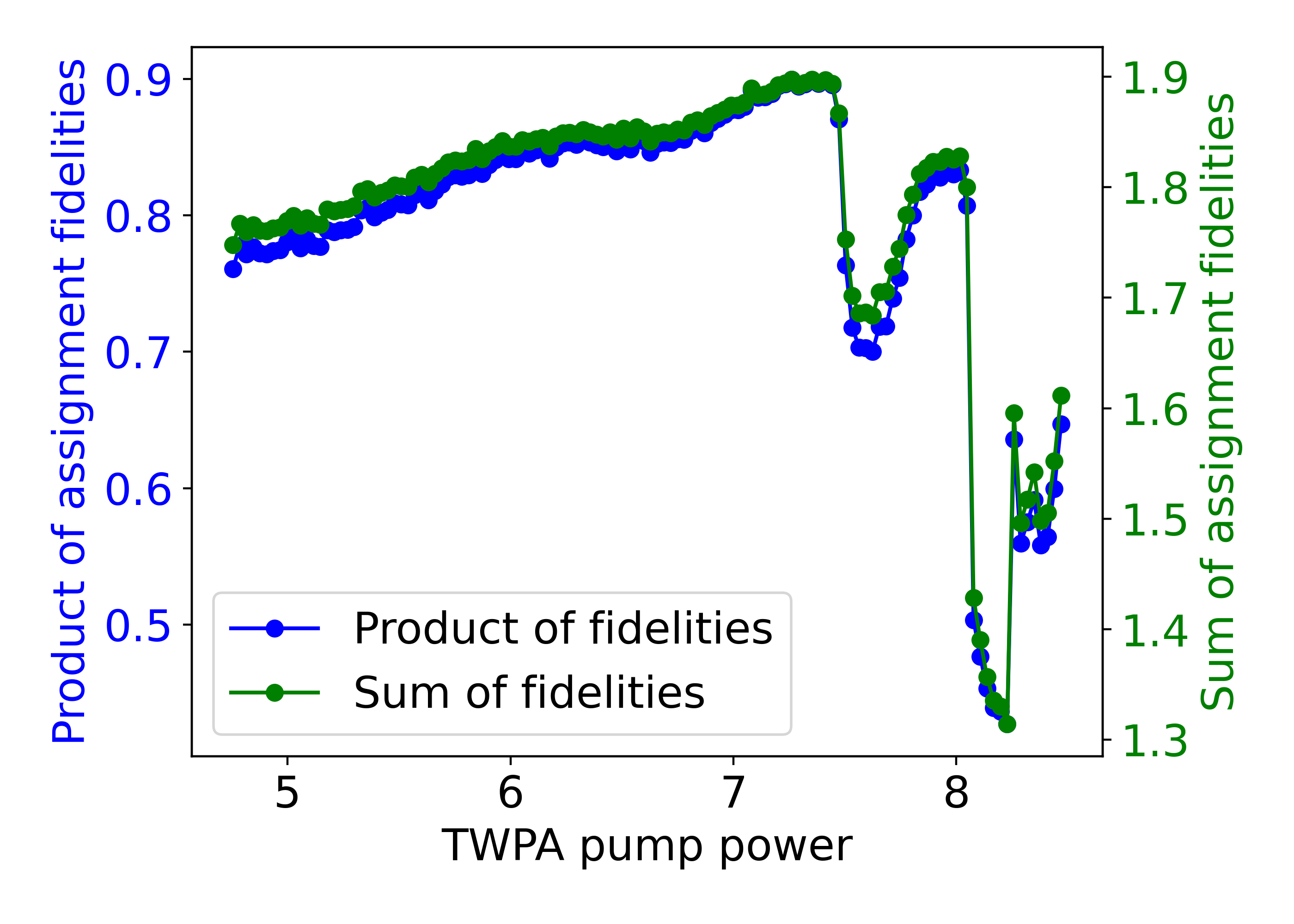}
    \caption{\justifying Sum and product of single-qubit assignment fidelities in System A. A sharp drop in both variables occurs at approximately $7.3\,\text{dBm}$ pump power.}
    \label{fig:twpa wider range}
\end{figure}

Pumping the parametric amplifier with increasing amplitude saturates the amplification and the signal quality drops off sharply at around $7.3\,\text{dBm}$ pump power (see Fig.~\ref{fig:twpa wider range}). This measurement was conducted with a different set of microwave cables, therefore there is a constant offset in power compared to Fig.~\ref{fig:inf threshold vs twpa power}.

\section{Theoretical shot budgeting}
\label{app:theory_shot_budgeting}
In a standard setup of detector tomography followed by state tomography, the distribution of shots should be dictated by the number of parameters to estimate in each part. If there is no knowledge of the quantum states to be estimated, then there are $4^{n}-1$ parameters for each normalized $n$-qubit state. An $m$-outcome POVM has $(m-1)4^n$ parameters that need to be estimated. If the POVM is informationally complete (IC), $m = 4^n$. Inserting this into eq.~\eqref{eq:ratio definition},
\begin{equation}
    r_n = \frac{(m-1)4^n}{(m-1)4^n + N_s(4^{n}-1)} \stackrel{\text{IC}}{=} \frac{4^n}{4^n + N_s},
\end{equation}
where $N_s$ is the number of states to be estimated with the same reconstructed POVM.
Compared with the experiment in Sec.~\ref{subsec:shot_budget}, setting $N_s = 1$, we get $r_1 = 0.80$ and $r_2 = 0.94$, significantly higher. 

The discrepancy between these two results may arise from the fact that not all the parameters are free and independent in a physical setting. For example, if the reconstructed states are close to pure, there are effectively $2(2^n-1)$ parameters in the state (since most parameters are close to zero). In addition, the POVM implemented in the experiment is not an IC POVM, but a set of projective measurements on the eigenstates of the Pauli operators, each individually normalized. 

If the errors at readout are close to classical (i.e. only redistributing the eigenvalues of the POVM elements), the number of parameters in the POVM is $(4^n - 1) 2^n$. If we assume close-to-classical errors, the ratio is
\begin{equation}
\begin{split}
     r_{n} &\stackrel{\text{mixed}}{=} \frac{2^n}{2^n + N_s},\\ 
     & \stackrel{\text{pure}}{=} \frac{4^n + 2^n}{4^n + 2^n + 2N_s} 
\end{split}
\end{equation}
which, compared to the experiment $(N_s = 1)$, gives us $r_1 = (0.67,~0.75)$ and $r_2 = (0.80,~0.91)$ for (mixed, pure) target states.

\section{Measurement drift}
\label{sec:drift}
\begin{figure*}[h!]
    \includegraphics[width=.48\linewidth]{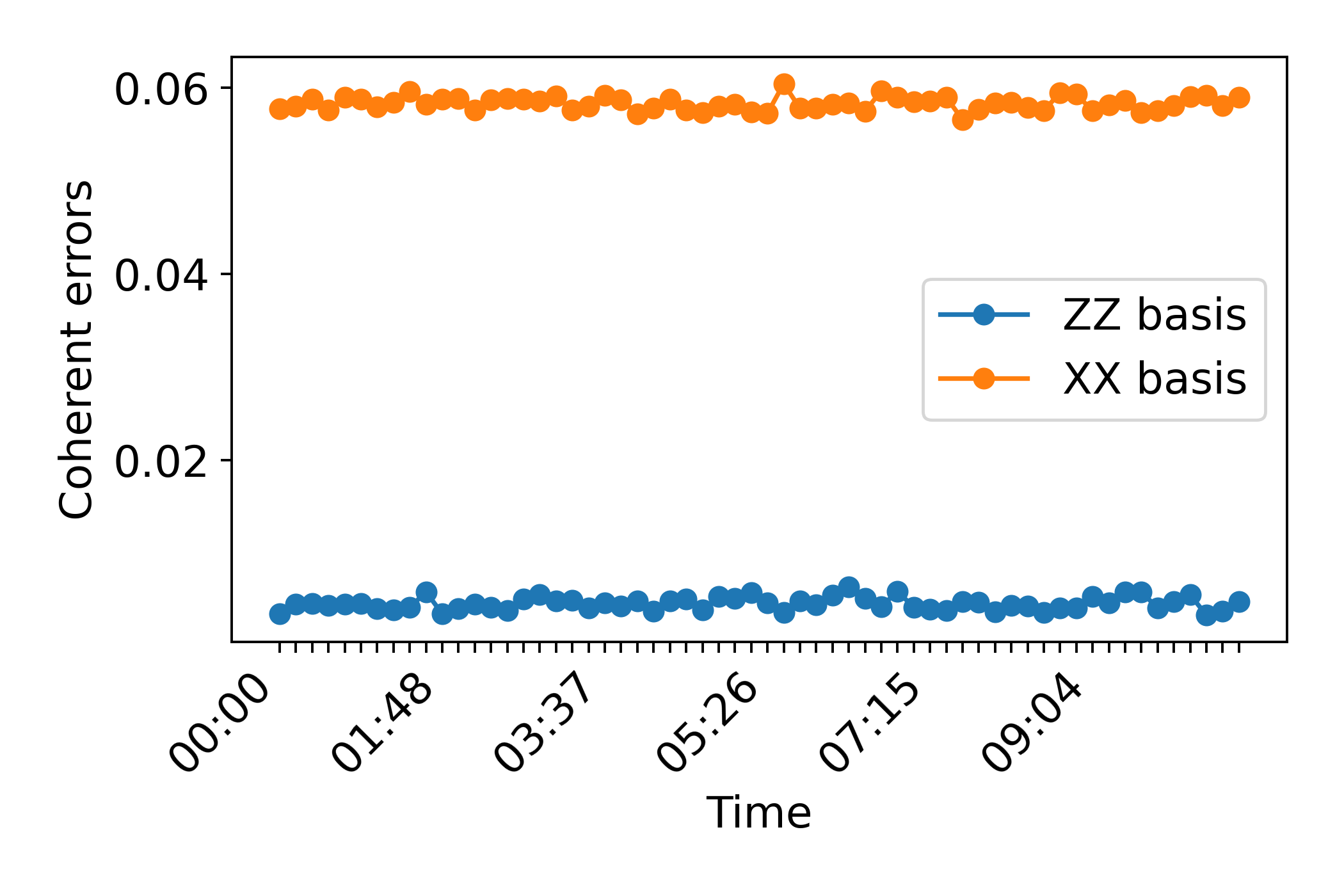}
    \includegraphics[width=.48\linewidth]{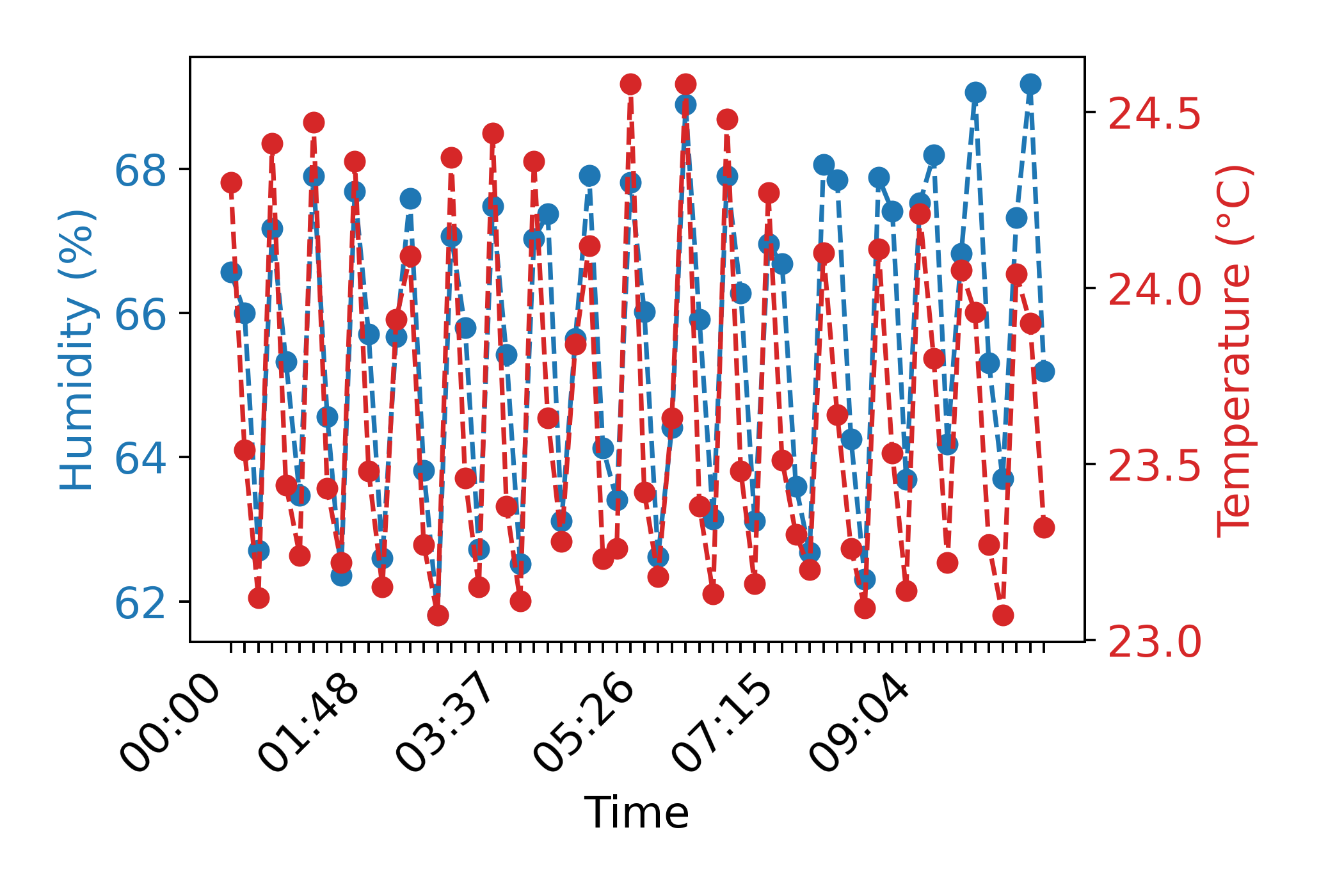}
    \includegraphics[width=.48\linewidth]{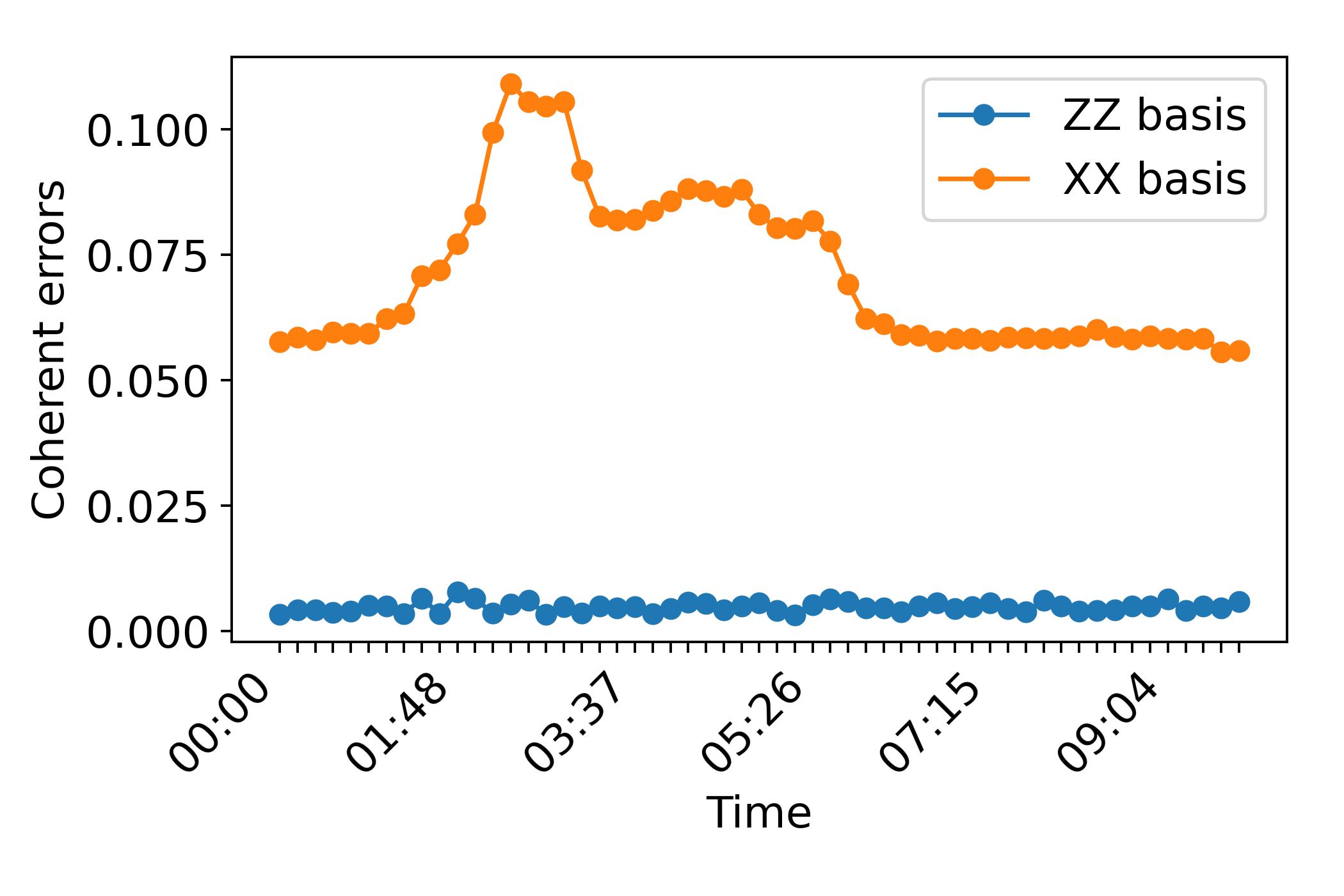}
    \includegraphics[width=.48\linewidth]{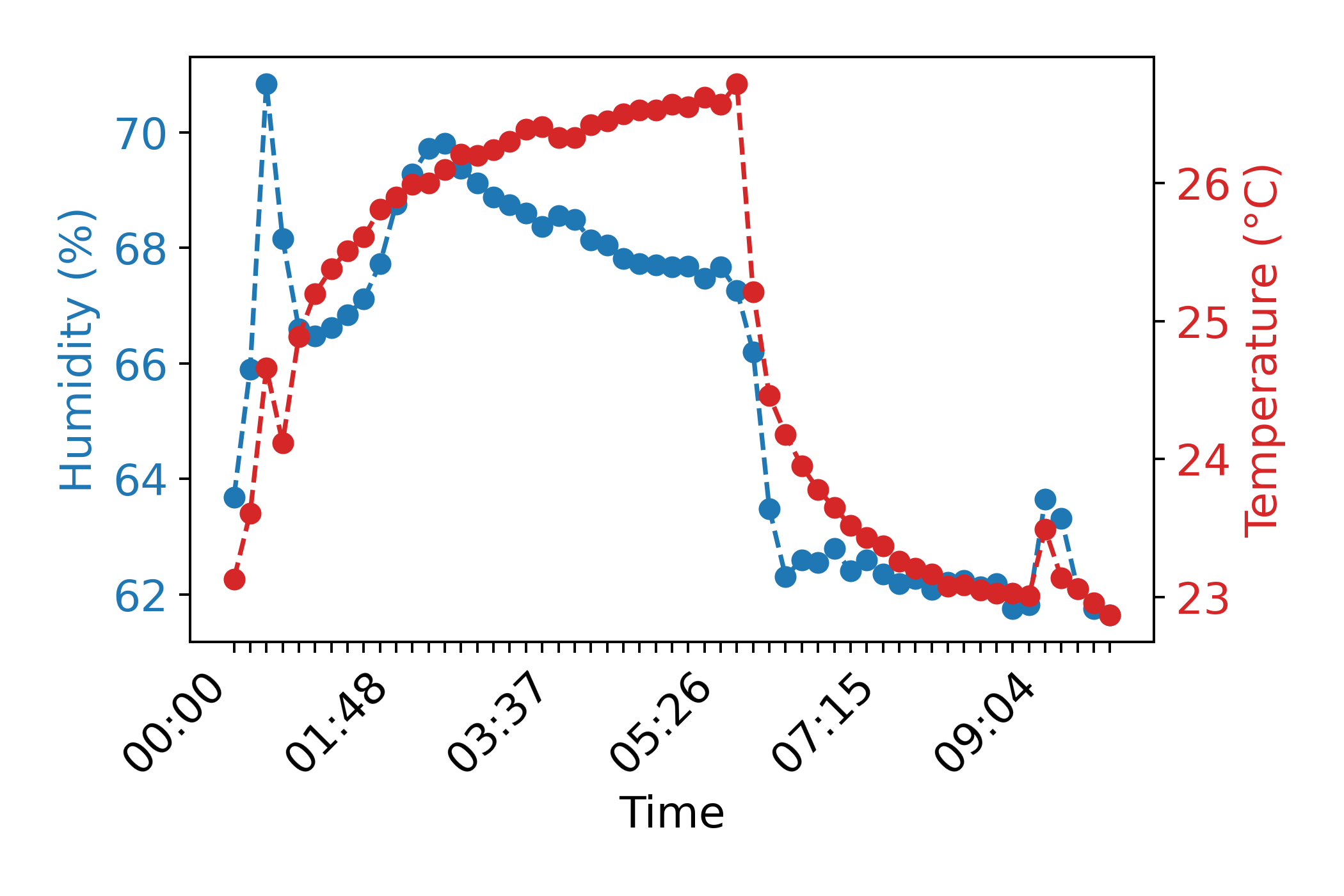}
    
    \caption{\justifying Stability of coherent readout errors with varied laboratory temperature and humidity. \textbf{Top left:} Coherent errors (defined in Eq.~\ref{eq:coh-err}) in the reconstructed measurement operators for $ZZ$- and $XX$-basis readout under stable laboratory conditions. \textbf{Top right:} Temperature and humidity values during the corresponding measurement. The rapid oscillations come from the feedback mechanism of the air conditioning. \textbf{Bottom left:} Coherent errors in the reconstructed measurement operators for $ZZ$- and $XX$-basis readout under unstable laboratory conditions induced by turning the air conditioning off, then on. \textbf{Bottom right:} Temperature and humidity values during the corresponding measurements.}
    \label{fig:povm-vs-time-drift}
\end{figure*}

As both QDT and QST scale exponentially in run time with the number of qubits, the experiment has to run self-consistently over a prolonged period. One limitation placed on the length of experiments is the temporal drift of the operational parameters. To quantitatively estimate the timescales and effects of these drifts on our POVMs, we repeatedly perform two-qubit detector tomography over 11 hours, while tracking the temperature and humidity of the laboratory.
Specifically, we reconstruct the measurement operators corresponding to $ZZ$ and $XX$ basis multiplexed readout of System A and plot their coherent errors in Fig.~\ref{fig:povm-vs-time-drift}. We define coherent errors in App.~\ref{app:coherent_errors}. While the coherent errors in the $ZZ$-basis are small and can be explained by statistical errors, the $XX$ basis contains significant coherent errors. This can be explained by the additional rotation gates used, which are not present in $ZZ$-basis readout. Under stable laboratory conditions, errors are observed to be temporally constant for both. Introducing rapid temperature and humidity changes by turning the laboratory air conditioning off, the amount of coherence in the reconstructed measurement operator of the $XX$ basis drifts significantly. As the $ZZ$-basis coherences are stable, we identify the likely source of coherence drift as originating from the manipulation of the qubit, used for $XX$-basis measurements.

\section{Three-qubit quantum state tomography with readout error mitigation}
\label{App: 3q qst}
\begin{figure*}[h!]
    \includegraphics[width=1\linewidth]{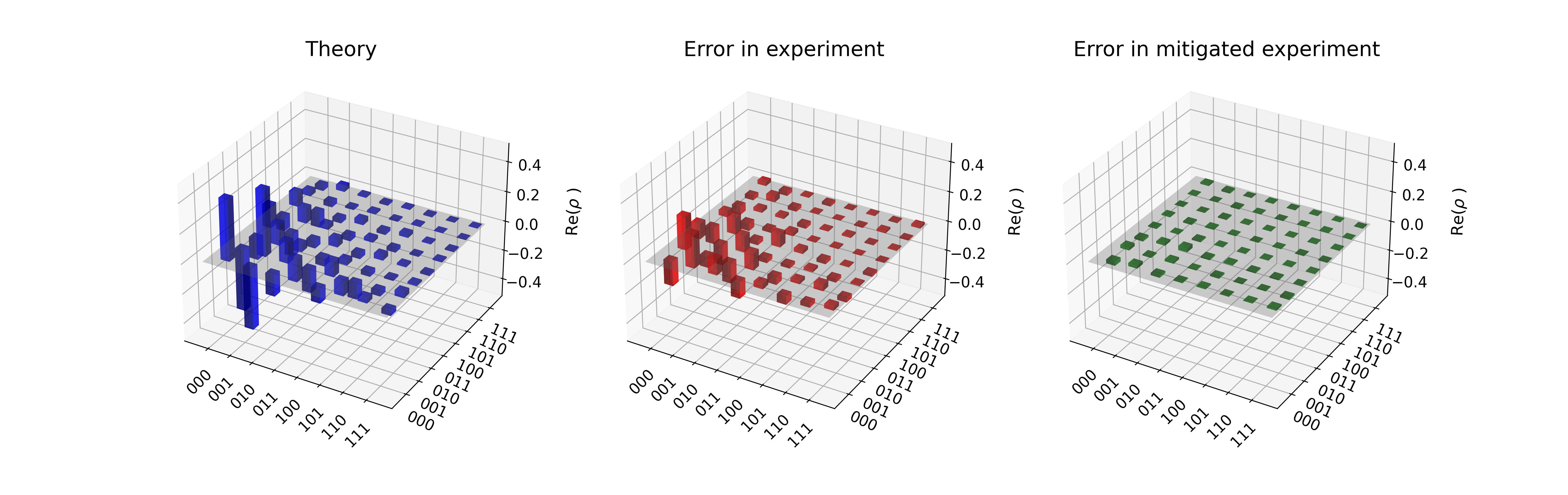}    \includegraphics[width=1\linewidth]{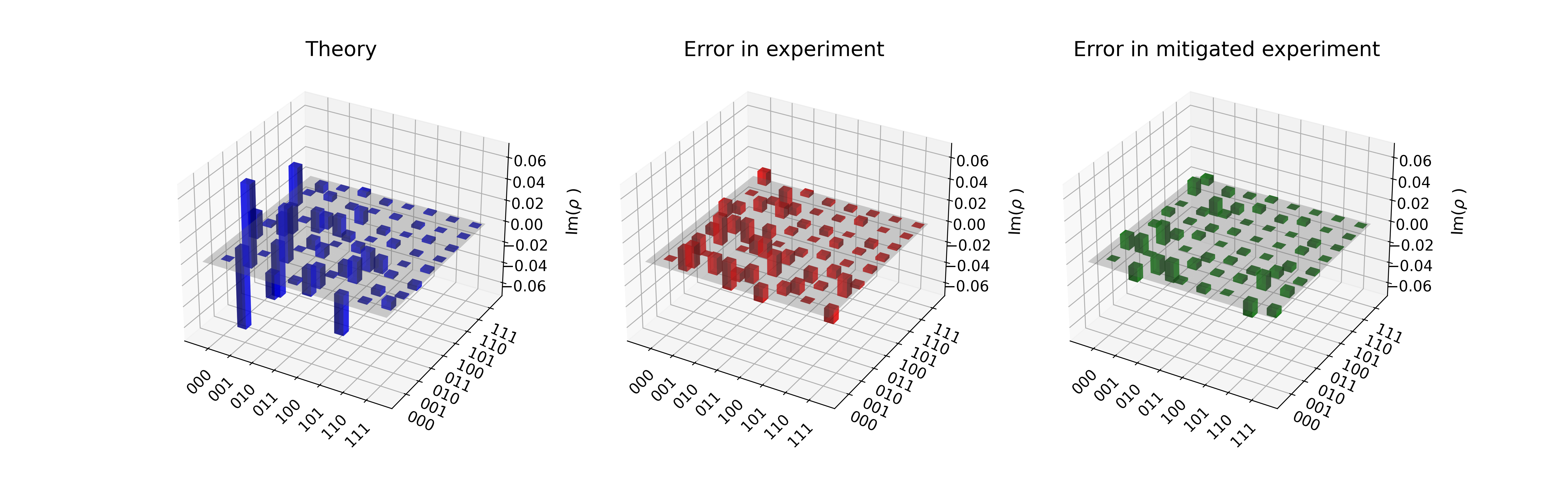}
    \caption{\justifying Comparison of theoretical density matrix of a three-qubit state and its reconstruction with and without readout-error mitigation. The top row shows the real part of the complex eight-dimensional matrices, and the bottom row shows the imaginary part. The quantum states reconstructed with and without readout-error mitigation have infidelities of 0.030 and 0.447, respectively. The different scales of the real and imaginary parts are due to the chosen quantum state. The errors in the imaginary and real parts of the error-mitigated experiments are of the same order of magnitude.}
    \label{app:3q-density-matrix}
\end{figure*}

The three-qubit density matrix obtained with readout-error mitigation has around an order-of-magnitude smaller error in the worst entry of the unmitigated density matrix (see Fig. \ref{app:3q-density-matrix}). The nonvanishing matrix entries have the largest errors both with and without readout-error mitigation. The infidelity of reconstruction was calculated from these density matrices.

\section{Coherent errors}
\label{app:coherent_errors}
One way to quantify the degree of nonclassicality of readout errors is to measure the distance between the reconstructed POVM and the nearest POVM subject to only classical errors.
Classical errors are captured by a confusion matrix \cite{Maciejewski2020}, which involves a statistical redistribution of the probabilities of measurement outcomes in the computational basis. Specifically, it can be described by a matrix $C$, which gives the relation between an ideal measurement probability $p_i$ of outcome $i$ and the measured frequencies $\tilde p_i = n_i/N$, 
\begin{equation}
    \tilde p_i = C_{ij}p_j.
    \label{eq:classical_errors}
\end{equation} 
Eq.~\eqref{eq:classical_errors} can be rephrased in terms of a noisy POVM $ \boldsymbol{\tilde M}$ and an ideal projective measurement $\boldsymbol{P}$ 
\begin{equation}
    \tilde M_i = C_{ij}P_i
    \label{eq:classical_POVMs}
\end{equation}
If coherent errors are present, then eq.~\eqref{eq:classical_POVMs} gets an additional term $\Delta_i$,
\begin{equation}
   \tilde M_i = C_{ij}P_i + \Delta_i.
    \label{eq:beyond_classical_errors}
\end{equation} 
Since every projective measurement operator  $P_i$ contains only diagonal entries, we can quantify coherent errors by computing the distance between the reconstructed POVM $\boldsymbol{\tilde M}$ and the same POVM with only diagonal entries \cite{Maciejewski2020},

\begin{equation}
    \epsilon_\text{coh} = \text{dist}(\boldsymbol{\tilde M},\mathcal{E}_\text{dep}(\boldsymbol{\tilde M})),
    \label{eq:coh-err}
\end{equation}
where $\text{dist}(.,.)$ is a suitable POVM distance, and $\mathcal{E}_\text{dep}(\tilde M_i) = \text{diag}(\tilde M_i)$ is the maximally dephasing channel. For coherent errors, we use the average-case distance for POVMs, which is defined as \cite{Maciejewski2023}
\begin{equation}
   d_{AV}(\boldsymbol{M}, \boldsymbol{N}) = \frac{1}{2d} \sum_{i=1}^n \sqrt{\norm{M_i - N_i}_{\text{HS}}^2+ \Tr(M_i - N_i)^2},
\end{equation}
where $n$ is the number of outcomes for the POVMs $\boldsymbol{M}$ and $\boldsymbol{N}$, $d$ is the dimension of the measurement operators, and $\norm{A}_\text{HS} = \sqrt{\Tr(A^2)}$.

\section{Effect of depolarizing noise on correlation coefficients}
\label{app:corr_coeff}
Many common sources of experimental errors, such as reduced computational-state distinguishability, can be modeled as a depolarizing channel on individual qubits. The single-qubit depolarizing channel can be written as \cite{Nielsen2012} 
\begin{equation}
    \mathcal{E}_\text{depol}(M_i) = \left(1-\frac{3k}{4}\right)M_i + \frac{k}{4}\left(XM_iX + YM_iY + ZM_iZ\right),
\end{equation}
where $k$ denotes the strength of depolarization and $X, Y$, and $Z$ are the Pauli matrices.

In the experiment discussed in Sec.~\ref{sec:correlation_under_state_distinguishability_noise}, the distinguishability of the two-qubit computational basis affects each qubit independently. We therefore expect the noise to be separable,
\begin{equation}
    \mathcal{E}^2_\text{depol}(M_i) = (\mathcal{E}_{\text{depol}} \otimes  \mathcal{E}_{\text{depol}}) (M_i).
    \label{eq:separable_depol_channel}
\end{equation}
Applying the separable depolarizing channel to the two-qubit measurement operators in the definition of the correlation coefficients, we find that
\begin{equation}
\begin{split}
    M_{x_i, \text{depol}}^{i, \sigma_j}=& \sum_{x_j} \Tr(  \mathcal{E}^2_{\text{depol}} (M_{x_i x_j}^{ij})(\mathrm{1}\otimes \sigma_j))\\
    =&  \sum_{x_j} \Tr(M_{x_i x_j}^{ij}(\mathbb{1}\otimes \mathcal{E}_\text{depol}(\sigma_j)))\\
    =&M_{x_i}^{i,\mathcal{E}_\text{depol}(\sigma_j)}.
\end{split}
\end{equation}
The correlation coefficients in eq.~\eqref{eq:quantum_correlation_coefficients} can therefore be written

\begin{equation}
        c_{j\rightarrow i} = \sup_{\rho_j, \sigma_j} ||(M^{i, \mathcal{E}_{\text{depol}}(\rho_j)}_0 - M^{i, \mathcal{E}_{\text{depol}}(\sigma_j)}_0)||_2.
\end{equation}
The depolarizing channel acts on the optimization states $\rho_j$ and $\sigma_j$ and effectively reduces the search space for the optimization. Consequently, the correlation coefficients are nonincreasing under single-qubit depolarizing noise.

\clearpage
\bibliography{refs}{}

 \newcommand{\noop}[1]{}
\begin{thebibliography}{46}%
\makeatletter
\providecommand \@ifxundefined [1]{%
 \@ifx{#1\undefined}
}%
\providecommand \@ifnum [1]{%
 \ifnum #1\expandafter \@firstoftwo
 \else \expandafter \@secondoftwo
 \fi
}%
\providecommand \@ifx [1]{%
 \ifx #1\expandafter \@firstoftwo
 \else \expandafter \@secondoftwo
 \fi
}%
\providecommand \natexlab [1]{#1}%
\providecommand \enquote  [1]{``#1''}%
\providecommand \bibnamefont  [1]{#1}%
\providecommand \bibfnamefont [1]{#1}%
\providecommand \citenamefont [1]{#1}%
\providecommand \href@noop [0]{\@secondoftwo}%
\providecommand \href [0]{\begingroup \@sanitize@url \@href}%
\providecommand \@href[1]{\@@startlink{#1}\@@href}%
\providecommand \@@href[1]{\endgroup#1\@@endlink}%
\providecommand \@sanitize@url [0]{\catcode `\\12\catcode `\$12\catcode `\&12\catcode `\#12\catcode `\^12\catcode `\_12\catcode `\%12\relax}%
\providecommand \@@startlink[1]{}%
\providecommand \@@endlink[0]{}%
\providecommand \url  [0]{\begingroup\@sanitize@url \@url }%
\providecommand \@url [1]{\endgroup\@href {#1}{\urlprefix }}%
\providecommand \urlprefix  [0]{URL }%
\providecommand \Eprint [0]{\href }%
\providecommand \doibase [0]{https://doi.org/}%
\providecommand \selectlanguage [0]{\@gobble}%
\providecommand \bibinfo  [0]{\@secondoftwo}%
\providecommand \bibfield  [0]{\@secondoftwo}%
\providecommand \translation [1]{[#1]}%
\providecommand \BibitemOpen [0]{}%
\providecommand \bibitemStop [0]{}%
\providecommand \bibitemNoStop [0]{.\EOS\space}%
\providecommand \EOS [0]{\spacefactor3000\relax}%
\providecommand \BibitemShut  [1]{\csname bibitem#1\endcsname}%
\let\auto@bib@innerbib\@empty
\bibitem [{\citenamefont {Acharya\mbox{, et. al.}}(2023)}]{Google2023}%
  \BibitemOpen
  \bibfield  {author} {\bibinfo {author} {\bibfnamefont {R.}~\bibnamefont {Acharya\mbox{, et. al.}}},\ }\bibfield  {title} {\bibinfo {title} {Suppressing quantum errors by scaling a surface code logical qubit},\ }\href {https://doi.org/10.1038/s41586-022-05434-1} {\bibfield  {journal} {\bibinfo  {journal} {Nature}\ }\textbf {\bibinfo {volume} {614}},\ \bibinfo {pages} {676–681} (\bibinfo {year} {2023})}\BibitemShut {NoStop}%
\bibitem [{\citenamefont {Acharya\mbox{, et. al.}}(2024)}]{Google2024}%
  \BibitemOpen
  \bibfield  {author} {\bibinfo {author} {\bibfnamefont {R.}~\bibnamefont {Acharya\mbox{, et. al.}}},\ }\bibfield  {title} {\bibinfo {title} {Quantum error correction below the surface code threshold},\ }\href {https://doi.org/10.1038/s41586-024-08449-y} {\bibfield  {journal} {\bibinfo  {journal} {Nature}\ }\textbf {\bibinfo {volume} {638}},\ \bibinfo {pages} {920–926} (\bibinfo {year} {2024})}\BibitemShut {NoStop}%
\bibitem [{\citenamefont {Li}\ and\ \citenamefont {Benjamin}(2017)}]{Li2017}%
  \BibitemOpen
  \bibfield  {author} {\bibinfo {author} {\bibfnamefont {Y.}~\bibnamefont {Li}}\ and\ \bibinfo {author} {\bibfnamefont {S.~C.}\ \bibnamefont {Benjamin}},\ }\bibfield  {title} {\bibinfo {title} {Efficient variational quantum simulator incorporating active error minimization},\ }\bibfield  {journal} {\bibinfo  {journal} {Physical Review X}\ }\textbf {\bibinfo {volume} {7}},\ \href {https://doi.org/10.1103/physrevx.7.021050} {10.1103/physrevx.7.021050} (\bibinfo {year} {2017})\BibitemShut {NoStop}%
\bibitem [{\citenamefont {Temme}\ \emph {et~al.}(2017)\citenamefont {Temme}, \citenamefont {Bravyi},\ and\ \citenamefont {Gambetta}}]{Temme2017}%
  \BibitemOpen
  \bibfield  {author} {\bibinfo {author} {\bibfnamefont {K.}~\bibnamefont {Temme}}, \bibinfo {author} {\bibfnamefont {S.}~\bibnamefont {Bravyi}},\ and\ \bibinfo {author} {\bibfnamefont {J.~M.}\ \bibnamefont {Gambetta}},\ }\bibfield  {title} {\bibinfo {title} {Error mitigation for short-depth quantum circuits},\ }\bibfield  {journal} {\bibinfo  {journal} {Physical Review Letters}\ }\textbf {\bibinfo {volume} {119}},\ \href {https://doi.org/10.1103/physrevlett.119.180509} {10.1103/physrevlett.119.180509} (\bibinfo {year} {2017})\BibitemShut {NoStop}%
\bibitem [{\citenamefont {Quek}\ \emph {et~al.}(2024)\citenamefont {Quek}, \citenamefont {Stilck~Fran\c{c}a}, \citenamefont {Khatri}, \citenamefont {Meyer},\ and\ \citenamefont {Eisert}}]{Quek2024}%
  \BibitemOpen
  \bibfield  {author} {\bibinfo {author} {\bibfnamefont {Y.}~\bibnamefont {Quek}}, \bibinfo {author} {\bibfnamefont {D.}~\bibnamefont {Stilck~Fran\c{c}a}}, \bibinfo {author} {\bibfnamefont {S.}~\bibnamefont {Khatri}}, \bibinfo {author} {\bibfnamefont {J.~J.}\ \bibnamefont {Meyer}},\ and\ \bibinfo {author} {\bibfnamefont {J.}~\bibnamefont {Eisert}},\ }\bibfield  {title} {\bibinfo {title} {Exponentially tighter bounds on limitations of quantum error mitigation},\ }\bibfield  {journal} {\bibinfo  {journal} {Nature Physics}\ }\href {https://doi.org/10.1038/s41567-024-02536-7} {10.1038/s41567-024-02536-7} (\bibinfo {year} {2024})\BibitemShut {NoStop}%
\bibitem [{\citenamefont {Li}\ \emph {et~al.}(2023)\citenamefont {Li}, \citenamefont {Liu}, \citenamefont {Zhao}, \citenamefont {Mi}, \citenamefont {Xu}, \citenamefont {Liang}, \citenamefont {Su}, \citenamefont {Sun}, \citenamefont {Xue}, \citenamefont {Zhang}, \citenamefont {Liu}, \citenamefont {Jin},\ and\ \citenamefont {Yu}}]{Li2023}%
  \BibitemOpen
  \bibfield  {author} {\bibinfo {author} {\bibfnamefont {Z.}~\bibnamefont {Li}}, \bibinfo {author} {\bibfnamefont {P.}~\bibnamefont {Liu}}, \bibinfo {author} {\bibfnamefont {P.}~\bibnamefont {Zhao}}, \bibinfo {author} {\bibfnamefont {Z.}~\bibnamefont {Mi}}, \bibinfo {author} {\bibfnamefont {H.}~\bibnamefont {Xu}}, \bibinfo {author} {\bibfnamefont {X.}~\bibnamefont {Liang}}, \bibinfo {author} {\bibfnamefont {T.}~\bibnamefont {Su}}, \bibinfo {author} {\bibfnamefont {W.}~\bibnamefont {Sun}}, \bibinfo {author} {\bibfnamefont {G.}~\bibnamefont {Xue}}, \bibinfo {author} {\bibfnamefont {J.-N.}\ \bibnamefont {Zhang}}, \bibinfo {author} {\bibfnamefont {W.}~\bibnamefont {Liu}}, \bibinfo {author} {\bibfnamefont {Y.}~\bibnamefont {Jin}},\ and\ \bibinfo {author} {\bibfnamefont {H.}~\bibnamefont {Yu}},\ }\bibfield  {title} {\bibinfo {title} {Error per single-qubit gate below 10-4 in a superconducting qubit},\ }\bibfield  {journal} {\bibinfo  {journal} {npj Quantum Information}\ }\textbf {\bibinfo {volume} {9}},\ \href
  {https://doi.org/10.1038/s41534-023-00781-x} {10.1038/s41534-023-00781-x} (\bibinfo {year} {2023})\BibitemShut {NoStop}%
\bibitem [{\citenamefont {Yan}\ \emph {et~al.}(2018)\citenamefont {Yan}, \citenamefont {Krantz}, \citenamefont {Sung}, \citenamefont {Kjaergaard}, \citenamefont {Campbell}, \citenamefont {Orlando}, \citenamefont {Gustavsson},\ and\ \citenamefont {Oliver}}]{Yan2018}%
  \BibitemOpen
  \bibfield  {author} {\bibinfo {author} {\bibfnamefont {F.}~\bibnamefont {Yan}}, \bibinfo {author} {\bibfnamefont {P.}~\bibnamefont {Krantz}}, \bibinfo {author} {\bibfnamefont {Y.}~\bibnamefont {Sung}}, \bibinfo {author} {\bibfnamefont {M.}~\bibnamefont {Kjaergaard}}, \bibinfo {author} {\bibfnamefont {D.~L.}\ \bibnamefont {Campbell}}, \bibinfo {author} {\bibfnamefont {T.~P.}\ \bibnamefont {Orlando}}, \bibinfo {author} {\bibfnamefont {S.}~\bibnamefont {Gustavsson}},\ and\ \bibinfo {author} {\bibfnamefont {W.~D.}\ \bibnamefont {Oliver}},\ }\bibfield  {title} {\bibinfo {title} {Tunable coupling scheme for implementing high-fidelity two-qubit gates},\ }\bibfield  {journal} {\bibinfo  {journal} {Physical Review Applied}\ }\textbf {\bibinfo {volume} {10}},\ \href {https://doi.org/10.1103/physrevapplied.10.054062} {10.1103/physrevapplied.10.054062} (\bibinfo {year} {2018})\BibitemShut {NoStop}%
\bibitem [{\citenamefont {Werninghaus}\ \emph {et~al.}(2021)\citenamefont {Werninghaus}, \citenamefont {Egger}, \citenamefont {Roy}, \citenamefont {Machnes}, \citenamefont {Wilhelm},\ and\ \citenamefont {Filipp}}]{Werninghaus2021}%
  \BibitemOpen
  \bibfield  {author} {\bibinfo {author} {\bibfnamefont {M.}~\bibnamefont {Werninghaus}}, \bibinfo {author} {\bibfnamefont {D.~J.}\ \bibnamefont {Egger}}, \bibinfo {author} {\bibfnamefont {F.}~\bibnamefont {Roy}}, \bibinfo {author} {\bibfnamefont {S.}~\bibnamefont {Machnes}}, \bibinfo {author} {\bibfnamefont {F.~K.}\ \bibnamefont {Wilhelm}},\ and\ \bibinfo {author} {\bibfnamefont {S.}~\bibnamefont {Filipp}},\ }\bibfield  {title} {\bibinfo {title} {Leakage reduction in fast superconducting qubit gates via optimal control},\ }\bibfield  {journal} {\bibinfo  {journal} {npj Quantum Information}\ }\textbf {\bibinfo {volume} {7}},\ \href {https://doi.org/10.1038/s41534-020-00346-2} {10.1038/s41534-020-00346-2} (\bibinfo {year} {2021})\BibitemShut {NoStop}%
\bibitem [{\citenamefont {Hyypp\"{a}}\ \emph {et~al.}(2024)\citenamefont {Hyypp\"{a}}, \citenamefont {Veps\"{a}l\"{a}inen}, \citenamefont {Papič}, \citenamefont {Chan}, \citenamefont {Inel}, \citenamefont {Landra}, \citenamefont {Liu}, \citenamefont {Luus}, \citenamefont {Marxer}, \citenamefont {Ockeloen-Korppi}, \citenamefont {Orbell}, \citenamefont {Tarasinski},\ and\ \citenamefont {Heinsoo}}]{Hyypp2024}%
  \BibitemOpen
  \bibfield  {author} {\bibinfo {author} {\bibfnamefont {E.}~\bibnamefont {Hyypp\"{a}}}, \bibinfo {author} {\bibfnamefont {A.}~\bibnamefont {Veps\"{a}l\"{a}inen}}, \bibinfo {author} {\bibfnamefont {M.}~\bibnamefont {Papič}}, \bibinfo {author} {\bibfnamefont {C.~F.}\ \bibnamefont {Chan}}, \bibinfo {author} {\bibfnamefont {S.}~\bibnamefont {Inel}}, \bibinfo {author} {\bibfnamefont {A.}~\bibnamefont {Landra}}, \bibinfo {author} {\bibfnamefont {W.}~\bibnamefont {Liu}}, \bibinfo {author} {\bibfnamefont {J.}~\bibnamefont {Luus}}, \bibinfo {author} {\bibfnamefont {F.}~\bibnamefont {Marxer}}, \bibinfo {author} {\bibfnamefont {C.}~\bibnamefont {Ockeloen-Korppi}}, \bibinfo {author} {\bibfnamefont {S.}~\bibnamefont {Orbell}}, \bibinfo {author} {\bibfnamefont {B.}~\bibnamefont {Tarasinski}},\ and\ \bibinfo {author} {\bibfnamefont {J.}~\bibnamefont {Heinsoo}},\ }\bibfield  {title} {\bibinfo {title} {Reducing leakage of single-qubit gates for superconducting quantum processors using analytical control pulse envelopes},\
  }\bibfield  {journal} {\bibinfo  {journal} {PRX Quantum}\ }\textbf {\bibinfo {volume} {5}},\ \href {https://doi.org/10.1103/prxquantum.5.030353} {10.1103/prxquantum.5.030353} (\bibinfo {year} {2024})\BibitemShut {NoStop}%
\bibitem [{\citenamefont {Marton}\ and\ \citenamefont {Asboth}(2023)}]{Mrton2023}%
  \BibitemOpen
  \bibfield  {author} {\bibinfo {author} {\bibfnamefont {A.}~\bibnamefont {Marton}}\ and\ \bibinfo {author} {\bibfnamefont {J.~K.}\ \bibnamefont {Asboth}},\ }\bibfield  {title} {\bibinfo {title} {Coherent errors and readout errors in the surface code},\ }\href {https://doi.org/10.22331/q-2023-09-21-1116} {\bibfield  {journal} {\bibinfo  {journal} {Quantum}\ }\textbf {\bibinfo {volume} {7}},\ \bibinfo {pages} {1116} (\bibinfo {year} {2023})}\BibitemShut {NoStop}%
\bibitem [{\citenamefont {Wilen}\ \emph {et~al.}(2021)\citenamefont {Wilen}, \citenamefont {Abdullah}, \citenamefont {Kurinsky}, \citenamefont {Stanford}, \citenamefont {Cardani}, \citenamefont {D’Imperio}, \citenamefont {Tomei}, \citenamefont {Faoro}, \citenamefont {Ioffe}, \citenamefont {Liu}, \citenamefont {Opremcak}, \citenamefont {Christensen}, \citenamefont {DuBois},\ and\ \citenamefont {McDermott}}]{Wilen2021}%
  \BibitemOpen
  \bibfield  {author} {\bibinfo {author} {\bibfnamefont {C.~D.}\ \bibnamefont {Wilen}}, \bibinfo {author} {\bibfnamefont {S.}~\bibnamefont {Abdullah}}, \bibinfo {author} {\bibfnamefont {N.~A.}\ \bibnamefont {Kurinsky}}, \bibinfo {author} {\bibfnamefont {C.}~\bibnamefont {Stanford}}, \bibinfo {author} {\bibfnamefont {L.}~\bibnamefont {Cardani}}, \bibinfo {author} {\bibfnamefont {G.}~\bibnamefont {D’Imperio}}, \bibinfo {author} {\bibfnamefont {C.}~\bibnamefont {Tomei}}, \bibinfo {author} {\bibfnamefont {L.}~\bibnamefont {Faoro}}, \bibinfo {author} {\bibfnamefont {L.~B.}\ \bibnamefont {Ioffe}}, \bibinfo {author} {\bibfnamefont {C.~H.}\ \bibnamefont {Liu}}, \bibinfo {author} {\bibfnamefont {A.}~\bibnamefont {Opremcak}}, \bibinfo {author} {\bibfnamefont {B.~G.}\ \bibnamefont {Christensen}}, \bibinfo {author} {\bibfnamefont {J.~L.}\ \bibnamefont {DuBois}},\ and\ \bibinfo {author} {\bibfnamefont {R.}~\bibnamefont {McDermott}},\ }\bibfield  {title} {\bibinfo {title} {Correlated charge noise and relaxation errors in
  superconducting qubits},\ }\href {https://doi.org/10.1038/s41586-021-03557-5} {\bibfield  {journal} {\bibinfo  {journal} {Nature}\ }\textbf {\bibinfo {volume} {594}},\ \bibinfo {pages} {369–373} (\bibinfo {year} {2021})}\BibitemShut {NoStop}%
\bibitem [{\citenamefont {Tuziemski}\ \emph {et~al.}(2023)\citenamefont {Tuziemski}, \citenamefont {Maciejewski}, \citenamefont {Majsak}, \citenamefont {Słowik}, \citenamefont {Kotowski}, \citenamefont {Kowalczyk-Murynka}, \citenamefont {Podziemski},\ and\ \citenamefont {Oszmaniec}}]{Tuziemski2023}%
  \BibitemOpen
  \bibfield  {author} {\bibinfo {author} {\bibfnamefont {J.}~\bibnamefont {Tuziemski}}, \bibinfo {author} {\bibfnamefont {F.~B.}\ \bibnamefont {Maciejewski}}, \bibinfo {author} {\bibfnamefont {J.}~\bibnamefont {Majsak}}, \bibinfo {author} {\bibfnamefont {O.}~\bibnamefont {Słowik}}, \bibinfo {author} {\bibfnamefont {M.}~\bibnamefont {Kotowski}}, \bibinfo {author} {\bibfnamefont {K.}~\bibnamefont {Kowalczyk-Murynka}}, \bibinfo {author} {\bibfnamefont {P.}~\bibnamefont {Podziemski}},\ and\ \bibinfo {author} {\bibfnamefont {M.}~\bibnamefont {Oszmaniec}},\ }\href {https://doi.org/10.48550/ARXIV.2311.10661} {\bibinfo {title} {Efficient reconstruction, benchmarking and validation of cross-talk models in readout noise in near-term quantum devices}} (\bibinfo {year} {2023})\BibitemShut {NoStop}%
\bibitem [{\citenamefont {Walter}\ \emph {et~al.}(2017)\citenamefont {Walter}, \citenamefont {Kurpiers}, \citenamefont {Gasparinetti}, \citenamefont {Magnard}, \citenamefont {Potočnik}, \citenamefont {Salathé}, \citenamefont {Pechal}, \citenamefont {Mondal}, \citenamefont {Oppliger}, \citenamefont {Eichler},\ and\ \citenamefont {Wallraff}}]{Walter2017}%
  \BibitemOpen
  \bibfield  {author} {\bibinfo {author} {\bibfnamefont {T.}~\bibnamefont {Walter}}, \bibinfo {author} {\bibfnamefont {P.}~\bibnamefont {Kurpiers}}, \bibinfo {author} {\bibfnamefont {S.}~\bibnamefont {Gasparinetti}}, \bibinfo {author} {\bibfnamefont {P.}~\bibnamefont {Magnard}}, \bibinfo {author} {\bibfnamefont {A.}~\bibnamefont {Potočnik}}, \bibinfo {author} {\bibfnamefont {Y.}~\bibnamefont {Salathé}}, \bibinfo {author} {\bibfnamefont {M.}~\bibnamefont {Pechal}}, \bibinfo {author} {\bibfnamefont {M.}~\bibnamefont {Mondal}}, \bibinfo {author} {\bibfnamefont {M.}~\bibnamefont {Oppliger}}, \bibinfo {author} {\bibfnamefont {C.}~\bibnamefont {Eichler}},\ and\ \bibinfo {author} {\bibfnamefont {A.}~\bibnamefont {Wallraff}},\ }\bibfield  {title} {\bibinfo {title} {Rapid high-fidelity single-shot dispersive readout of superconducting qubits},\ }\bibfield  {journal} {\bibinfo  {journal} {Physical Review Applied}\ }\textbf {\bibinfo {volume} {7}},\ \href {https://doi.org/10.1103/physrevapplied.7.054020}
  {10.1103/physrevapplied.7.054020} (\bibinfo {year} {2017})\BibitemShut {NoStop}%
\bibitem [{\citenamefont {Heinsoo}\ \emph {et~al.}(2018)\citenamefont {Heinsoo}, \citenamefont {Andersen}, \citenamefont {Remm}, \citenamefont {Krinner}, \citenamefont {Walter}, \citenamefont {Salathé}, \citenamefont {Gasparinetti}, \citenamefont {Besse}, \citenamefont {Potočnik}, \citenamefont {Wallraff},\ and\ \citenamefont {Eichler}}]{Heinsoo2018}%
  \BibitemOpen
  \bibfield  {author} {\bibinfo {author} {\bibfnamefont {J.}~\bibnamefont {Heinsoo}}, \bibinfo {author} {\bibfnamefont {C.~K.}\ \bibnamefont {Andersen}}, \bibinfo {author} {\bibfnamefont {A.}~\bibnamefont {Remm}}, \bibinfo {author} {\bibfnamefont {S.}~\bibnamefont {Krinner}}, \bibinfo {author} {\bibfnamefont {T.}~\bibnamefont {Walter}}, \bibinfo {author} {\bibfnamefont {Y.}~\bibnamefont {Salathé}}, \bibinfo {author} {\bibfnamefont {S.}~\bibnamefont {Gasparinetti}}, \bibinfo {author} {\bibfnamefont {J.-C.}\ \bibnamefont {Besse}}, \bibinfo {author} {\bibfnamefont {A.}~\bibnamefont {Potočnik}}, \bibinfo {author} {\bibfnamefont {A.}~\bibnamefont {Wallraff}},\ and\ \bibinfo {author} {\bibfnamefont {C.}~\bibnamefont {Eichler}},\ }\bibfield  {title} {\bibinfo {title} {Rapid high-fidelity multiplexed readout of superconducting qubits},\ }\bibfield  {journal} {\bibinfo  {journal} {Physical Review Applied}\ }\textbf {\bibinfo {volume} {10}},\ \href {https://doi.org/10.1103/physrevapplied.10.034040}
  {10.1103/physrevapplied.10.034040} (\bibinfo {year} {2018})\BibitemShut {NoStop}%
\bibitem [{\citenamefont {Lundeen}\ \emph {et~al.}(2008)\citenamefont {Lundeen}, \citenamefont {Feito}, \citenamefont {Coldenstrodt-Ronge}, \citenamefont {Pregnell}, \citenamefont {Silberhorn}, \citenamefont {Ralph}, \citenamefont {Eisert}, \citenamefont {Plenio},\ and\ \citenamefont {Walmsley}}]{Lundeen2008}%
  \BibitemOpen
  \bibfield  {author} {\bibinfo {author} {\bibfnamefont {J.~S.}\ \bibnamefont {Lundeen}}, \bibinfo {author} {\bibfnamefont {A.}~\bibnamefont {Feito}}, \bibinfo {author} {\bibfnamefont {H.}~\bibnamefont {Coldenstrodt-Ronge}}, \bibinfo {author} {\bibfnamefont {K.~L.}\ \bibnamefont {Pregnell}}, \bibinfo {author} {\bibfnamefont {C.}~\bibnamefont {Silberhorn}}, \bibinfo {author} {\bibfnamefont {T.~C.}\ \bibnamefont {Ralph}}, \bibinfo {author} {\bibfnamefont {J.}~\bibnamefont {Eisert}}, \bibinfo {author} {\bibfnamefont {M.~B.}\ \bibnamefont {Plenio}},\ and\ \bibinfo {author} {\bibfnamefont {I.~A.}\ \bibnamefont {Walmsley}},\ }\bibfield  {title} {\bibinfo {title} {Tomography of quantum detectors},\ }\href {https://doi.org/10.1038/nphys1133} {\bibfield  {journal} {\bibinfo  {journal} {Nature Physics}\ }\textbf {\bibinfo {volume} {5}},\ \bibinfo {pages} {27–30} (\bibinfo {year} {2008})}\BibitemShut {NoStop}%
\bibitem [{\citenamefont {Chen}\ \emph {et~al.}(2019)\citenamefont {Chen}, \citenamefont {Farahzad}, \citenamefont {Yoo},\ and\ \citenamefont {Wei}}]{Chen2019}%
  \BibitemOpen
  \bibfield  {author} {\bibinfo {author} {\bibfnamefont {Y.}~\bibnamefont {Chen}}, \bibinfo {author} {\bibfnamefont {M.}~\bibnamefont {Farahzad}}, \bibinfo {author} {\bibfnamefont {S.}~\bibnamefont {Yoo}},\ and\ \bibinfo {author} {\bibfnamefont {T.-C.}\ \bibnamefont {Wei}},\ }\bibfield  {title} {\bibinfo {title} {Detector tomography on {IBM} quantum computers and mitigation of an imperfect measurement},\ }\bibfield  {journal} {\bibinfo  {journal} {Physical Review A}\ }\textbf {\bibinfo {volume} {100}},\ \href {https://doi.org/10.1103/physreva.100.052315} {10.1103/physreva.100.052315} (\bibinfo {year} {2019})\BibitemShut {NoStop}%
\bibitem [{\citenamefont {Maciejewski}\ \emph {et~al.}(2020)\citenamefont {Maciejewski}, \citenamefont {Zimbor{\'{a}}s},\ and\ \citenamefont {Oszmaniec}}]{Maciejewski2020}%
  \BibitemOpen
  \bibfield  {author} {\bibinfo {author} {\bibfnamefont {F.~B.}\ \bibnamefont {Maciejewski}}, \bibinfo {author} {\bibfnamefont {Z.}~\bibnamefont {Zimbor{\'{a}}s}},\ and\ \bibinfo {author} {\bibfnamefont {M.}~\bibnamefont {Oszmaniec}},\ }\bibfield  {title} {\bibinfo {title} {Mitigation of readout noise in near-term quantum devices by classical post-processing based on detector tomography},\ }\href {https://doi.org/10.22331/q-2020-04-24-257} {\bibfield  {journal} {\bibinfo  {journal} {Quantum}\ }\textbf {\bibinfo {volume} {4}},\ \bibinfo {pages} {257} (\bibinfo {year} {2020})}\BibitemShut {NoStop}%
\bibitem [{\citenamefont {Maciejewski}\ \emph {et~al.}(2021)\citenamefont {Maciejewski}, \citenamefont {Baccari}, \citenamefont {Zimborás},\ and\ \citenamefont {Oszmaniec}}]{Maciejewski2021}%
  \BibitemOpen
  \bibfield  {author} {\bibinfo {author} {\bibfnamefont {F.~B.}\ \bibnamefont {Maciejewski}}, \bibinfo {author} {\bibfnamefont {F.}~\bibnamefont {Baccari}}, \bibinfo {author} {\bibfnamefont {Z.}~\bibnamefont {Zimborás}},\ and\ \bibinfo {author} {\bibfnamefont {M.}~\bibnamefont {Oszmaniec}},\ }\bibfield  {title} {\bibinfo {title} {Modeling and mitigation of cross-talk effects in readout noise with applications to the quantum approximate optimization algorithm},\ }\href {https://doi.org/10.22331/q-2021-06-01-464} {\bibfield  {journal} {\bibinfo  {journal} {Quantum}\ }\textbf {\bibinfo {volume} {5}},\ \bibinfo {pages} {464} (\bibinfo {year} {2021})}\BibitemShut {NoStop}%
\bibitem [{\citenamefont {Kono}\ \emph {et~al.}(2024)\citenamefont {Kono}, \citenamefont {Pan}, \citenamefont {Chegnizadeh}, \citenamefont {Wang}, \citenamefont {Youssefi}, \citenamefont {Scigliuzzo},\ and\ \citenamefont {Kippenberg}}]{Kono2024}%
  \BibitemOpen
  \bibfield  {author} {\bibinfo {author} {\bibfnamefont {S.}~\bibnamefont {Kono}}, \bibinfo {author} {\bibfnamefont {J.}~\bibnamefont {Pan}}, \bibinfo {author} {\bibfnamefont {M.}~\bibnamefont {Chegnizadeh}}, \bibinfo {author} {\bibfnamefont {X.}~\bibnamefont {Wang}}, \bibinfo {author} {\bibfnamefont {A.}~\bibnamefont {Youssefi}}, \bibinfo {author} {\bibfnamefont {M.}~\bibnamefont {Scigliuzzo}},\ and\ \bibinfo {author} {\bibfnamefont {T.~J.}\ \bibnamefont {Kippenberg}},\ }\bibfield  {title} {\bibinfo {title} {Mechanically induced correlated errors on superconducting qubits with relaxation times exceeding 0.4 ms},\ }\bibfield  {journal} {\bibinfo  {journal} {Nature Communications}\ }\textbf {\bibinfo {volume} {15}},\ \href {https://doi.org/10.1038/s41467-024-48230-3} {10.1038/s41467-024-48230-3} (\bibinfo {year} {2024})\BibitemShut {NoStop}%
\bibitem [{\citenamefont {Li}\ \emph {et~al.}(2025)\citenamefont {Li}, \citenamefont {Wang}, \citenamefont {Jiang}, \citenamefont {Xue}, \citenamefont {Cai}, \citenamefont {Zhou}, \citenamefont {Gong}, \citenamefont {Liu}, \citenamefont {Zheng}, \citenamefont {Ma}, \citenamefont {Chen}, \citenamefont {Sun}, \citenamefont {Yang}, \citenamefont {Yan}, \citenamefont {Jin}, \citenamefont {Zhao}, \citenamefont {Ding},\ and\ \citenamefont {Yu}}]{Li2025}%
  \BibitemOpen
  \bibfield  {author} {\bibinfo {author} {\bibfnamefont {X.}~\bibnamefont {Li}}, \bibinfo {author} {\bibfnamefont {J.}~\bibnamefont {Wang}}, \bibinfo {author} {\bibfnamefont {Y.-Y.}\ \bibnamefont {Jiang}}, \bibinfo {author} {\bibfnamefont {G.-M.}\ \bibnamefont {Xue}}, \bibinfo {author} {\bibfnamefont {X.}~\bibnamefont {Cai}}, \bibinfo {author} {\bibfnamefont {J.}~\bibnamefont {Zhou}}, \bibinfo {author} {\bibfnamefont {M.}~\bibnamefont {Gong}}, \bibinfo {author} {\bibfnamefont {Z.-F.}\ \bibnamefont {Liu}}, \bibinfo {author} {\bibfnamefont {S.-Y.}\ \bibnamefont {Zheng}}, \bibinfo {author} {\bibfnamefont {D.-K.}\ \bibnamefont {Ma}}, \bibinfo {author} {\bibfnamefont {M.}~\bibnamefont {Chen}}, \bibinfo {author} {\bibfnamefont {W.-J.}\ \bibnamefont {Sun}}, \bibinfo {author} {\bibfnamefont {S.}~\bibnamefont {Yang}}, \bibinfo {author} {\bibfnamefont {F.}~\bibnamefont {Yan}}, \bibinfo {author} {\bibfnamefont {Y.-R.}\ \bibnamefont {Jin}}, \bibinfo {author} {\bibfnamefont {S.~P.}\ \bibnamefont {Zhao}}, \bibinfo {author}
  {\bibfnamefont {X.-F.}\ \bibnamefont {Ding}},\ and\ \bibinfo {author} {\bibfnamefont {H.-F.}\ \bibnamefont {Yu}},\ }\bibfield  {title} {\bibinfo {title} {Cosmic-ray-induced correlated errors in superconducting qubit array},\ }\bibfield  {journal} {\bibinfo  {journal} {Nature Communications}\ }\textbf {\bibinfo {volume} {16}},\ \href {https://doi.org/10.1038/s41467-025-59778-z} {10.1038/s41467-025-59778-z} (\bibinfo {year} {2025})\BibitemShut {NoStop}%
\bibitem [{\citenamefont {Aasen}\ \emph {et~al.}(2025)\citenamefont {Aasen}, \citenamefont {Di~Giovanni}, \citenamefont {Rotzinger}, \citenamefont {Ustinov},\ and\ \citenamefont {G\"arttner}}]{Aasen2025}%
  \BibitemOpen
  \bibfield  {author} {\bibinfo {author} {\bibfnamefont {A.~S.}\ \bibnamefont {Aasen}}, \bibinfo {author} {\bibfnamefont {A.}~\bibnamefont {Di~Giovanni}}, \bibinfo {author} {\bibfnamefont {H.}~\bibnamefont {Rotzinger}}, \bibinfo {author} {\bibfnamefont {A.~V.}\ \bibnamefont {Ustinov}},\ and\ \bibinfo {author} {\bibfnamefont {M.}~\bibnamefont {G\"arttner}},\ }\bibfield  {title} {\bibinfo {title} {Mitigation of correlated readout errors without randomized measurements},\ }\href {https://doi.org/10.1103/6p6s-t8b7} {\bibfield  {journal} {\bibinfo  {journal} {Phys. Rev. A}\ }\textbf {\bibinfo {volume} {112}},\ \bibinfo {pages} {022601} (\bibinfo {year} {2025})}\BibitemShut {NoStop}%
\bibitem [{\citenamefont {de~Groot}\ \emph {et~al.}(2010)\citenamefont {de~Groot}, \citenamefont {van Loo}, \citenamefont {Lisenfeld}, \citenamefont {Schouten}, \citenamefont {Lupaşcu}, \citenamefont {Harmans},\ and\ \citenamefont {Mooij}}]{deGroot2010}%
  \BibitemOpen
  \bibfield  {author} {\bibinfo {author} {\bibfnamefont {P.~C.}\ \bibnamefont {de~Groot}}, \bibinfo {author} {\bibfnamefont {A.~F.}\ \bibnamefont {van Loo}}, \bibinfo {author} {\bibfnamefont {J.}~\bibnamefont {Lisenfeld}}, \bibinfo {author} {\bibfnamefont {R.~N.}\ \bibnamefont {Schouten}}, \bibinfo {author} {\bibfnamefont {A.}~\bibnamefont {Lupaşcu}}, \bibinfo {author} {\bibfnamefont {C.~J. P.~M.}\ \bibnamefont {Harmans}},\ and\ \bibinfo {author} {\bibfnamefont {J.~E.}\ \bibnamefont {Mooij}},\ }\bibfield  {title} {\bibinfo {title} {Low-crosstalk bifurcation detectors for coupled flux qubits},\ }\bibfield  {journal} {\bibinfo  {journal} {Applied Physics Letters}\ }\textbf {\bibinfo {volume} {96}},\ \href {https://doi.org/10.1063/1.3367875} {10.1063/1.3367875} (\bibinfo {year} {2010})\BibitemShut {NoStop}%
\bibitem [{\citenamefont {Aasen}\ \emph {et~al.}(2024)\citenamefont {Aasen}, \citenamefont {Di~Giovanni}, \citenamefont {Rotzinger}, \citenamefont {Ustinov},\ and\ \citenamefont {G\"{a}rttner}}]{Aasen2024}%
  \BibitemOpen
  \bibfield  {author} {\bibinfo {author} {\bibfnamefont {A.~S.}\ \bibnamefont {Aasen}}, \bibinfo {author} {\bibfnamefont {A.}~\bibnamefont {Di~Giovanni}}, \bibinfo {author} {\bibfnamefont {H.}~\bibnamefont {Rotzinger}}, \bibinfo {author} {\bibfnamefont {A.~V.}\ \bibnamefont {Ustinov}},\ and\ \bibinfo {author} {\bibfnamefont {M.}~\bibnamefont {G\"{a}rttner}},\ }\bibfield  {title} {\bibinfo {title} {Readout error mitigated quantum state tomography tested on superconducting qubits},\ }\bibfield  {journal} {\bibinfo  {journal} {Communications Physics}\ }\textbf {\bibinfo {volume} {7}},\ \href {https://doi.org/10.1038/s42005-024-01790-8} {10.1038/s42005-024-01790-8} (\bibinfo {year} {2024})\BibitemShut {NoStop}%
\bibitem [{\citenamefont {Geller}(2021)}]{Geller2021}%
  \BibitemOpen
  \bibfield  {author} {\bibinfo {author} {\bibfnamefont {M.~R.}\ \bibnamefont {Geller}},\ }\bibfield  {title} {\bibinfo {title} {Conditionally rigorous mitigation of multiqubit measurement errors},\ }\href {https://doi.org/10.1103/PhysRevLett.127.090502} {\bibfield  {journal} {\bibinfo  {journal} {Phys. Rev. Lett.}\ }\textbf {\bibinfo {volume} {127}},\ \bibinfo {pages} {090502} (\bibinfo {year} {2021})}\BibitemShut {NoStop}%
\bibitem [{\citenamefont {Geller}\ and\ \citenamefont {Sun}(2021)}]{Geller2021-b}%
  \BibitemOpen
  \bibfield  {author} {\bibinfo {author} {\bibfnamefont {M.~R.}\ \bibnamefont {Geller}}\ and\ \bibinfo {author} {\bibfnamefont {M.}~\bibnamefont {Sun}},\ }\bibfield  {title} {\bibinfo {title} {Toward efficient correction of multiqubit measurement errors: pair correlation method},\ }\href {https://doi.org/10.1088/2058-9565/abd5c9} {\bibfield  {journal} {\bibinfo  {journal} {Quantum Science and Technology}\ }\textbf {\bibinfo {volume} {6}},\ \bibinfo {pages} {025009} (\bibinfo {year} {2021})}\BibitemShut {NoStop}%
\bibitem [{\citenamefont {Jerger}\ \emph {et~al.}(2012)\citenamefont {Jerger}, \citenamefont {Poletto}, \citenamefont {Macha}, \citenamefont {H\"{u}bner}, \citenamefont {Il’ichev},\ and\ \citenamefont {Ustinov}}]{Jerger2012}%
  \BibitemOpen
  \bibfield  {author} {\bibinfo {author} {\bibfnamefont {M.}~\bibnamefont {Jerger}}, \bibinfo {author} {\bibfnamefont {S.}~\bibnamefont {Poletto}}, \bibinfo {author} {\bibfnamefont {P.}~\bibnamefont {Macha}}, \bibinfo {author} {\bibfnamefont {U.}~\bibnamefont {H\"{u}bner}}, \bibinfo {author} {\bibfnamefont {E.}~\bibnamefont {Il’ichev}},\ and\ \bibinfo {author} {\bibfnamefont {A.~V.}\ \bibnamefont {Ustinov}},\ }\bibfield  {title} {\bibinfo {title} {Frequency division multiplexing readout and simultaneous manipulation of an array of flux qubits},\ }\href {https://doi.org/10.1063/1.4739454} {\bibfield  {journal} {\bibinfo  {journal} {Applied Physics Letters}\ }\textbf {\bibinfo {volume} {101}},\ \bibinfo {pages} {042604} (\bibinfo {year} {2012})}\BibitemShut {NoStop}%
\bibitem [{\citenamefont {Macklin}\ \emph {et~al.}(2015)\citenamefont {Macklin}, \citenamefont {O’Brien}, \citenamefont {Hover}, \citenamefont {Schwartz}, \citenamefont {Bolkhovsky}, \citenamefont {Zhang}, \citenamefont {Oliver},\ and\ \citenamefont {Siddiqi}}]{Macklin2015}%
  \BibitemOpen
  \bibfield  {author} {\bibinfo {author} {\bibfnamefont {C.}~\bibnamefont {Macklin}}, \bibinfo {author} {\bibfnamefont {K.}~\bibnamefont {O’Brien}}, \bibinfo {author} {\bibfnamefont {D.}~\bibnamefont {Hover}}, \bibinfo {author} {\bibfnamefont {M.~E.}\ \bibnamefont {Schwartz}}, \bibinfo {author} {\bibfnamefont {V.}~\bibnamefont {Bolkhovsky}}, \bibinfo {author} {\bibfnamefont {X.}~\bibnamefont {Zhang}}, \bibinfo {author} {\bibfnamefont {W.~D.}\ \bibnamefont {Oliver}},\ and\ \bibinfo {author} {\bibfnamefont {I.}~\bibnamefont {Siddiqi}},\ }\bibfield  {title} {\bibinfo {title} {A near–quantum-limited josephson traveling-wave parametric amplifier},\ }\href {https://doi.org/10.1126/science.aaa8525} {\bibfield  {journal} {\bibinfo  {journal} {Science}\ }\textbf {\bibinfo {volume} {350}},\ \bibinfo {pages} {307–310} (\bibinfo {year} {2015})}\BibitemShut {NoStop}%
\bibitem [{\citenamefont {Kim}\ \emph {et~al.}(2022)\citenamefont {Kim}, \citenamefont {Morvan}, \citenamefont {Nguyen}, \citenamefont {Naik}, \citenamefont {J\"{u}nger}, \citenamefont {Chen}, \citenamefont {Kreikebaum}, \citenamefont {Santiago},\ and\ \citenamefont {Siddiqi}}]{Kim2022}%
  \BibitemOpen
  \bibfield  {author} {\bibinfo {author} {\bibfnamefont {Y.}~\bibnamefont {Kim}}, \bibinfo {author} {\bibfnamefont {A.}~\bibnamefont {Morvan}}, \bibinfo {author} {\bibfnamefont {L.~B.}\ \bibnamefont {Nguyen}}, \bibinfo {author} {\bibfnamefont {R.~K.}\ \bibnamefont {Naik}}, \bibinfo {author} {\bibfnamefont {C.}~\bibnamefont {J\"{u}nger}}, \bibinfo {author} {\bibfnamefont {L.}~\bibnamefont {Chen}}, \bibinfo {author} {\bibfnamefont {J.~M.}\ \bibnamefont {Kreikebaum}}, \bibinfo {author} {\bibfnamefont {D.~I.}\ \bibnamefont {Santiago}},\ and\ \bibinfo {author} {\bibfnamefont {I.}~\bibnamefont {Siddiqi}},\ }\bibfield  {title} {\bibinfo {title} {High-fidelity three-qubit itoffoli gate for fixed-frequency superconducting qubits},\ }\href {https://doi.org/10.1038/s41567-022-01590-3} {\bibfield  {journal} {\bibinfo  {journal} {Nature Physics}\ }\textbf {\bibinfo {volume} {18}},\ \bibinfo {pages} {783–788} (\bibinfo {year} {2022})}\BibitemShut {NoStop}%
\bibitem [{\citenamefont {Itoko}\ \emph {et~al.}(2024)\citenamefont {Itoko}, \citenamefont {Malekakhlagh}, \citenamefont {Kanazawa},\ and\ \citenamefont {Takita}}]{Itoko2024}%
  \BibitemOpen
  \bibfield  {author} {\bibinfo {author} {\bibfnamefont {T.}~\bibnamefont {Itoko}}, \bibinfo {author} {\bibfnamefont {M.}~\bibnamefont {Malekakhlagh}}, \bibinfo {author} {\bibfnamefont {N.}~\bibnamefont {Kanazawa}},\ and\ \bibinfo {author} {\bibfnamefont {M.}~\bibnamefont {Takita}},\ }\bibfield  {title} {\bibinfo {title} {Three-qubit parity gate via simultaneous cross-resonance drives},\ }\bibfield  {journal} {\bibinfo  {journal} {Physical Review Applied}\ }\textbf {\bibinfo {volume} {21}},\ \href {https://doi.org/10.1103/physrevapplied.21.034018} {10.1103/physrevapplied.21.034018} (\bibinfo {year} {2024})\BibitemShut {NoStop}%
\bibitem [{\citenamefont {Li}\ \emph {et~al.}(2024)\citenamefont {Li}, \citenamefont {Tao}, \citenamefont {Yi}, \citenamefont {Luo}, \citenamefont {Zhang}, \citenamefont {Zhou}, \citenamefont {Liu}, \citenamefont {Yan}, \citenamefont {Chen},\ and\ \citenamefont {Yu}}]{Li2024}%
  \BibitemOpen
  \bibfield  {author} {\bibinfo {author} {\bibfnamefont {X.-L.}\ \bibnamefont {Li}}, \bibinfo {author} {\bibfnamefont {Z.}~\bibnamefont {Tao}}, \bibinfo {author} {\bibfnamefont {K.}~\bibnamefont {Yi}}, \bibinfo {author} {\bibfnamefont {K.}~\bibnamefont {Luo}}, \bibinfo {author} {\bibfnamefont {L.}~\bibnamefont {Zhang}}, \bibinfo {author} {\bibfnamefont {Y.}~\bibnamefont {Zhou}}, \bibinfo {author} {\bibfnamefont {S.}~\bibnamefont {Liu}}, \bibinfo {author} {\bibfnamefont {T.}~\bibnamefont {Yan}}, \bibinfo {author} {\bibfnamefont {Y.}~\bibnamefont {Chen}},\ and\ \bibinfo {author} {\bibfnamefont {D.}~\bibnamefont {Yu}},\ }\bibfield  {title} {\bibinfo {title} {Hardware-efficient and fast three-qubit gate in superconducting quantum circuits},\ }\bibfield  {journal} {\bibinfo  {journal} {Frontiers of Physics}\ }\textbf {\bibinfo {volume} {19}},\ \href {https://doi.org/10.1007/s11467-024-1405-8} {10.1007/s11467-024-1405-8} (\bibinfo {year} {2024})\BibitemShut {NoStop}%
\bibitem [{\citenamefont {Thorbeck}\ \emph {et~al.}(2023)\citenamefont {Thorbeck}, \citenamefont {Eddins}, \citenamefont {Lauer}, \citenamefont {McClure},\ and\ \citenamefont {Carroll}}]{Thorbeck2023}%
  \BibitemOpen
  \bibfield  {author} {\bibinfo {author} {\bibfnamefont {T.}~\bibnamefont {Thorbeck}}, \bibinfo {author} {\bibfnamefont {A.}~\bibnamefont {Eddins}}, \bibinfo {author} {\bibfnamefont {I.}~\bibnamefont {Lauer}}, \bibinfo {author} {\bibfnamefont {D.~T.}\ \bibnamefont {McClure}},\ and\ \bibinfo {author} {\bibfnamefont {M.}~\bibnamefont {Carroll}},\ }\bibfield  {title} {\bibinfo {title} {Two-level-system dynamics in a superconducting qubit due to background ionizing radiation},\ }\bibfield  {journal} {\bibinfo  {journal} {PRX Quantum}\ }\textbf {\bibinfo {volume} {4}},\ \href {https://doi.org/10.1103/prxquantum.4.020356} {10.1103/prxquantum.4.020356} (\bibinfo {year} {2023})\BibitemShut {NoStop}%
\bibitem [{\citenamefont {Fiur{\'{a}}{\v{s}}ek}(2001)}]{Fiurek2001}%
  \BibitemOpen
  \bibfield  {author} {\bibinfo {author} {\bibfnamefont {J.}~\bibnamefont {Fiur{\'{a}}{\v{s}}ek}},\ }\bibfield  {title} {\bibinfo {title} {Maximum-likelihood estimation of quantum measurement},\ }\bibfield  {journal} {\bibinfo  {journal} {Physical Review A}\ }\textbf {\bibinfo {volume} {64}},\ \href {https://doi.org/10.1103/physreva.64.024102} {10.1103/physreva.64.024102} (\bibinfo {year} {2001})\BibitemShut {NoStop}%
\bibitem [{\citenamefont {Lvovsky}(2004)}]{Lvovsky2004}%
  \BibitemOpen
  \bibfield  {author} {\bibinfo {author} {\bibfnamefont {A.~I.}\ \bibnamefont {Lvovsky}},\ }\bibfield  {title} {\bibinfo {title} {Iterative maximum-likelihood reconstruction in quantum homodyne tomography},\ }\href {https://doi.org/10.1088/1464-4266/6/6/014} {\bibfield  {journal} {\bibinfo  {journal} {Journal of Optics B: Quantum and Semiclassical Optics}\ }\textbf {\bibinfo {volume} {6}},\ \bibinfo {pages} {S556} (\bibinfo {year} {2004})}\BibitemShut {NoStop}%
\bibitem [{\citenamefont {Struchalin}\ \emph {et~al.}(2016)\citenamefont {Struchalin}, \citenamefont {Pogorelov}, \citenamefont {Straupe}, \citenamefont {Kravtsov}, \citenamefont {Radchenko},\ and\ \citenamefont {Kulik}}]{Struchalin2016}%
  \BibitemOpen
  \bibfield  {author} {\bibinfo {author} {\bibfnamefont {G.~I.}\ \bibnamefont {Struchalin}}, \bibinfo {author} {\bibfnamefont {I.~A.}\ \bibnamefont {Pogorelov}}, \bibinfo {author} {\bibfnamefont {S.~S.}\ \bibnamefont {Straupe}}, \bibinfo {author} {\bibfnamefont {K.~S.}\ \bibnamefont {Kravtsov}}, \bibinfo {author} {\bibfnamefont {I.~V.}\ \bibnamefont {Radchenko}},\ and\ \bibinfo {author} {\bibfnamefont {S.~P.}\ \bibnamefont {Kulik}},\ }\bibfield  {title} {\bibinfo {title} {Experimental adaptive quantum tomography of two-qubit states},\ }\bibfield  {journal} {\bibinfo  {journal} {Physical Review A}\ }\textbf {\bibinfo {volume} {93}},\ \href {https://doi.org/10.1103/physreva.93.012103} {10.1103/physreva.93.012103} (\bibinfo {year} {2016})\BibitemShut {NoStop}%
\bibitem [{\citenamefont {Lienhard}\ \emph {et~al.}(2022)\citenamefont {Lienhard}, \citenamefont {Veps\"al\"ainen}, \citenamefont {Govia}, \citenamefont {Hoffer}, \citenamefont {Qiu}, \citenamefont {Rist\`e}, \citenamefont {Ware}, \citenamefont {Kim}, \citenamefont {Winik}, \citenamefont {Melville}, \citenamefont {Niedzielski}, \citenamefont {Yoder}, \citenamefont {Ribeill}, \citenamefont {Ohki}, \citenamefont {Krovi}, \citenamefont {Orlando}, \citenamefont {Gustavsson},\ and\ \citenamefont {Oliver}}]{Lienhard2021}%
  \BibitemOpen
  \bibfield  {author} {\bibinfo {author} {\bibfnamefont {B.}~\bibnamefont {Lienhard}}, \bibinfo {author} {\bibfnamefont {A.}~\bibnamefont {Veps\"al\"ainen}}, \bibinfo {author} {\bibfnamefont {L.~C.}\ \bibnamefont {Govia}}, \bibinfo {author} {\bibfnamefont {C.~R.}\ \bibnamefont {Hoffer}}, \bibinfo {author} {\bibfnamefont {J.~Y.}\ \bibnamefont {Qiu}}, \bibinfo {author} {\bibfnamefont {D.}~\bibnamefont {Rist\`e}}, \bibinfo {author} {\bibfnamefont {M.}~\bibnamefont {Ware}}, \bibinfo {author} {\bibfnamefont {D.}~\bibnamefont {Kim}}, \bibinfo {author} {\bibfnamefont {R.}~\bibnamefont {Winik}}, \bibinfo {author} {\bibfnamefont {A.}~\bibnamefont {Melville}}, \bibinfo {author} {\bibfnamefont {B.}~\bibnamefont {Niedzielski}}, \bibinfo {author} {\bibfnamefont {J.}~\bibnamefont {Yoder}}, \bibinfo {author} {\bibfnamefont {G.~J.}\ \bibnamefont {Ribeill}}, \bibinfo {author} {\bibfnamefont {T.~A.}\ \bibnamefont {Ohki}}, \bibinfo {author} {\bibfnamefont {H.~K.}\ \bibnamefont {Krovi}}, \bibinfo {author} {\bibfnamefont {T.~P.}\
  \bibnamefont {Orlando}}, \bibinfo {author} {\bibfnamefont {S.}~\bibnamefont {Gustavsson}},\ and\ \bibinfo {author} {\bibfnamefont {W.~D.}\ \bibnamefont {Oliver}},\ }\bibfield  {title} {\bibinfo {title} {Deep-neural-network discrimination of multiplexed superconducting-qubit states},\ }\href {https://doi.org/10.1103/PhysRevApplied.17.014024} {\bibfield  {journal} {\bibinfo  {journal} {Phys. Rev. Appl.}\ }\textbf {\bibinfo {volume} {17}},\ \bibinfo {pages} {014024} (\bibinfo {year} {2022})}\BibitemShut {NoStop}%
\bibitem [{\citenamefont {Barends}\ \emph {et~al.}(2013)\citenamefont {Barends}, \citenamefont {Kelly}, \citenamefont {Megrant}, \citenamefont {Sank}, \citenamefont {Jeffrey}, \citenamefont {Chen}, \citenamefont {Yin}, \citenamefont {Chiaro}, \citenamefont {Mutus}, \citenamefont {Neill}, \citenamefont {O’Malley}, \citenamefont {Roushan}, \citenamefont {Wenner}, \citenamefont {White}, \citenamefont {Cleland},\ and\ \citenamefont {Martinis}}]{Barends2013}%
  \BibitemOpen
  \bibfield  {author} {\bibinfo {author} {\bibfnamefont {R.}~\bibnamefont {Barends}}, \bibinfo {author} {\bibfnamefont {J.}~\bibnamefont {Kelly}}, \bibinfo {author} {\bibfnamefont {A.}~\bibnamefont {Megrant}}, \bibinfo {author} {\bibfnamefont {D.}~\bibnamefont {Sank}}, \bibinfo {author} {\bibfnamefont {E.}~\bibnamefont {Jeffrey}}, \bibinfo {author} {\bibfnamefont {Y.}~\bibnamefont {Chen}}, \bibinfo {author} {\bibfnamefont {Y.}~\bibnamefont {Yin}}, \bibinfo {author} {\bibfnamefont {B.}~\bibnamefont {Chiaro}}, \bibinfo {author} {\bibfnamefont {J.}~\bibnamefont {Mutus}}, \bibinfo {author} {\bibfnamefont {C.}~\bibnamefont {Neill}}, \bibinfo {author} {\bibfnamefont {P.}~\bibnamefont {O’Malley}}, \bibinfo {author} {\bibfnamefont {P.}~\bibnamefont {Roushan}}, \bibinfo {author} {\bibfnamefont {J.}~\bibnamefont {Wenner}}, \bibinfo {author} {\bibfnamefont {T.~C.}\ \bibnamefont {White}}, \bibinfo {author} {\bibfnamefont {A.~N.}\ \bibnamefont {Cleland}},\ and\ \bibinfo {author} {\bibfnamefont {J.~M.}\ \bibnamefont
  {Martinis}},\ }\bibfield  {title} {\bibinfo {title} {Coherent josephson qubit suitable for scalable quantum integrated circuits},\ }\bibfield  {journal} {\bibinfo  {journal} {Physical Review Letters}\ }\textbf {\bibinfo {volume} {111}},\ \href {https://doi.org/10.1103/physrevlett.111.080502} {10.1103/physrevlett.111.080502} (\bibinfo {year} {2013})\BibitemShut {NoStop}%
\bibitem [{\citenamefont {Mezzadri}(2006)}]{mezzadri2006}%
  \BibitemOpen
  \bibfield  {author} {\bibinfo {author} {\bibfnamefont {F.}~\bibnamefont {Mezzadri}},\ }\href@noop {} {\bibinfo {title} {How to generate random matrices from the classical compact groups}} (\bibinfo {year} {2006}),\ \Eprint {https://arxiv.org/abs/math-ph/0609050} {arXiv:math-ph/0609050 [math-ph]} \BibitemShut {NoStop}%
\bibitem [{\citenamefont {Życzkowski}\ \emph {et~al.}(2011)\citenamefont {Życzkowski}, \citenamefont {Penson}, \citenamefont {Nechita},\ and\ \citenamefont {Collins}}]{yczkowski2011}%
  \BibitemOpen
  \bibfield  {author} {\bibinfo {author} {\bibfnamefont {K.}~\bibnamefont {Życzkowski}}, \bibinfo {author} {\bibfnamefont {K.~A.}\ \bibnamefont {Penson}}, \bibinfo {author} {\bibfnamefont {I.}~\bibnamefont {Nechita}},\ and\ \bibinfo {author} {\bibfnamefont {B.}~\bibnamefont {Collins}},\ }\bibfield  {title} {\bibinfo {title} {Generating random density matrices},\ }\bibfield  {journal} {\bibinfo  {journal} {Journal of Mathematical Physics}\ }\textbf {\bibinfo {volume} {52}},\ \href {https://doi.org/10.1063/1.3595693} {10.1063/1.3595693} (\bibinfo {year} {2011})\BibitemShut {NoStop}%
\bibitem [{\citenamefont {Warren}\ \emph {et~al.}(2023)\citenamefont {Warren}, \citenamefont {Fernández-Pendás}, \citenamefont {Ahmed}, \citenamefont {Abad}, \citenamefont {Bengtsson}, \citenamefont {Biznárová}, \citenamefont {Debnath}, \citenamefont {Gu}, \citenamefont {Križan}, \citenamefont {Osman}, \citenamefont {Fadavi~Roudsari}, \citenamefont {Delsing}, \citenamefont {Johansson}, \citenamefont {Frisk~Kockum}, \citenamefont {Tancredi},\ and\ \citenamefont {Bylander}}]{Warren2023}%
  \BibitemOpen
  \bibfield  {author} {\bibinfo {author} {\bibfnamefont {C.~W.}\ \bibnamefont {Warren}}, \bibinfo {author} {\bibfnamefont {J.}~\bibnamefont {Fernández-Pendás}}, \bibinfo {author} {\bibfnamefont {S.}~\bibnamefont {Ahmed}}, \bibinfo {author} {\bibfnamefont {T.}~\bibnamefont {Abad}}, \bibinfo {author} {\bibfnamefont {A.}~\bibnamefont {Bengtsson}}, \bibinfo {author} {\bibfnamefont {J.}~\bibnamefont {Biznárová}}, \bibinfo {author} {\bibfnamefont {K.}~\bibnamefont {Debnath}}, \bibinfo {author} {\bibfnamefont {X.}~\bibnamefont {Gu}}, \bibinfo {author} {\bibfnamefont {C.}~\bibnamefont {Križan}}, \bibinfo {author} {\bibfnamefont {A.}~\bibnamefont {Osman}}, \bibinfo {author} {\bibfnamefont {A.}~\bibnamefont {Fadavi~Roudsari}}, \bibinfo {author} {\bibfnamefont {P.}~\bibnamefont {Delsing}}, \bibinfo {author} {\bibfnamefont {G.}~\bibnamefont {Johansson}}, \bibinfo {author} {\bibfnamefont {A.}~\bibnamefont {Frisk~Kockum}}, \bibinfo {author} {\bibfnamefont {G.}~\bibnamefont {Tancredi}},\ and\ \bibinfo {author}
  {\bibfnamefont {J.}~\bibnamefont {Bylander}},\ }\bibfield  {title} {\bibinfo {title} {Extensive characterization and implementation of a family of three-qubit gates at the coherence limit},\ }\bibfield  {journal} {\bibinfo  {journal} {npj Quantum Information}\ }\textbf {\bibinfo {volume} {9}},\ \href {https://doi.org/10.1038/s41534-023-00711-x} {10.1038/s41534-023-00711-x} (\bibinfo {year} {2023})\BibitemShut {NoStop}%
\bibitem [{\citenamefont {Kreuzer}\ \emph {et~al.}(2025)\citenamefont {Kreuzer}, \citenamefont {Krumrey}, \citenamefont {Tohamy}, \citenamefont {Ionita}, \citenamefont {Rotzinger},\ and\ \citenamefont {Ustinov}}]{Kreuzer2025}%
  \BibitemOpen
  \bibfield  {author} {\bibinfo {author} {\bibfnamefont {A.}~\bibnamefont {Kreuzer}}, \bibinfo {author} {\bibfnamefont {T.}~\bibnamefont {Krumrey}}, \bibinfo {author} {\bibfnamefont {H.}~\bibnamefont {Tohamy}}, \bibinfo {author} {\bibfnamefont {A.}~\bibnamefont {Ionita}}, \bibinfo {author} {\bibfnamefont {H.}~\bibnamefont {Rotzinger}},\ and\ \bibinfo {author} {\bibfnamefont {A.~V.}\ \bibnamefont {Ustinov}},\ }\href {https://arxiv.org/abs/2503.11437} {\bibinfo {title} {Stacked josephson junctions for quantum circuit applications}} (\bibinfo {year} {2025}),\ \Eprint {https://arxiv.org/abs/2503.11437} {arXiv:2503.11437 [cond-mat.supr-con]} \BibitemShut {NoStop}%
\bibitem [{\citenamefont {Gunyhó}\ \emph {et~al.}(2024)\citenamefont {Gunyhó}, \citenamefont {Kundu}, \citenamefont {Ma}, \citenamefont {Liu}, \citenamefont {Niemel\"{a}}, \citenamefont {Catto}, \citenamefont {Vadimov}, \citenamefont {Vesterinen}, \citenamefont {Singh}, \citenamefont {Chen},\ and\ \citenamefont {M\"{o}tt\"{o}nen}}]{Gunyh2024}%
  \BibitemOpen
  \bibfield  {author} {\bibinfo {author} {\bibfnamefont {A.~M.}\ \bibnamefont {Gunyhó}}, \bibinfo {author} {\bibfnamefont {S.}~\bibnamefont {Kundu}}, \bibinfo {author} {\bibfnamefont {J.}~\bibnamefont {Ma}}, \bibinfo {author} {\bibfnamefont {W.}~\bibnamefont {Liu}}, \bibinfo {author} {\bibfnamefont {S.}~\bibnamefont {Niemel\"{a}}}, \bibinfo {author} {\bibfnamefont {G.}~\bibnamefont {Catto}}, \bibinfo {author} {\bibfnamefont {V.}~\bibnamefont {Vadimov}}, \bibinfo {author} {\bibfnamefont {V.}~\bibnamefont {Vesterinen}}, \bibinfo {author} {\bibfnamefont {P.}~\bibnamefont {Singh}}, \bibinfo {author} {\bibfnamefont {Q.}~\bibnamefont {Chen}},\ and\ \bibinfo {author} {\bibfnamefont {M.}~\bibnamefont {M\"{o}tt\"{o}nen}},\ }\bibfield  {title} {\bibinfo {title} {Single-shot readout of a superconducting qubit using a thermal detector},\ }\href {https://doi.org/10.1038/s41928-024-01147-7} {\bibfield  {journal} {\bibinfo  {journal} {Nature Electronics}\ }\textbf {\bibinfo {volume} {7}},\ \bibinfo {pages} {288–298}
  (\bibinfo {year} {2024})}\BibitemShut {NoStop}%
\bibitem [{\citenamefont {Aasen}(2025)}]{Aasen2025Github}%
  \BibitemOpen
  \bibfield  {author} {\bibinfo {author} {\bibfnamefont {A.}~\bibnamefont {Aasen}},\ }\href@noop {} {\bibinfo {title} {{CREMST}}},\ \bibinfo {howpublished} {\url{https://github.com/AdrianAasen/CREMST}} (\bibinfo {year} {2025})\BibitemShut {NoStop}%
\bibitem [{\citenamefont {Cotler}\ and\ \citenamefont {Wilczek}(2020)}]{Cotler2020}%
  \BibitemOpen
  \bibfield  {author} {\bibinfo {author} {\bibfnamefont {J.}~\bibnamefont {Cotler}}\ and\ \bibinfo {author} {\bibfnamefont {F.}~\bibnamefont {Wilczek}},\ }\bibfield  {title} {\bibinfo {title} {Quantum overlapping tomography},\ }\bibfield  {journal} {\bibinfo  {journal} {Physical Review Letters}\ }\textbf {\bibinfo {volume} {124}},\ \href {https://doi.org/10.1103/physrevlett.124.100401} {10.1103/physrevlett.124.100401} (\bibinfo {year} {2020})\BibitemShut {NoStop}%
\bibitem [{\citenamefont {Zhu}\ \emph {et~al.}(2013)\citenamefont {Zhu}, \citenamefont {Ferguson}, \citenamefont {Manucharyan},\ and\ \citenamefont {Koch}}]{Zhu2013}%
  \BibitemOpen
  \bibfield  {author} {\bibinfo {author} {\bibfnamefont {G.}~\bibnamefont {Zhu}}, \bibinfo {author} {\bibfnamefont {D.~G.}\ \bibnamefont {Ferguson}}, \bibinfo {author} {\bibfnamefont {V.~E.}\ \bibnamefont {Manucharyan}},\ and\ \bibinfo {author} {\bibfnamefont {J.}~\bibnamefont {Koch}},\ }\bibfield  {title} {\bibinfo {title} {Circuit qed with fluxonium qubits: Theory of the dispersive regime},\ }\bibfield  {journal} {\bibinfo  {journal} {Physical Review B}\ }\textbf {\bibinfo {volume} {87}},\ \href {https://doi.org/10.1103/physrevb.87.024510} {10.1103/physrevb.87.024510} (\bibinfo {year} {2013})\BibitemShut {NoStop}%
\bibitem [{\citenamefont {Maciejewski}\ \emph {et~al.}(2023)\citenamefont {Maciejewski}, \citenamefont {Puchała},\ and\ \citenamefont {Oszmaniec}}]{Maciejewski2023}%
  \BibitemOpen
  \bibfield  {author} {\bibinfo {author} {\bibfnamefont {F.~B.}\ \bibnamefont {Maciejewski}}, \bibinfo {author} {\bibfnamefont {Z.}~\bibnamefont {Puchała}},\ and\ \bibinfo {author} {\bibfnamefont {M.}~\bibnamefont {Oszmaniec}},\ }\bibfield  {title} {\bibinfo {title} {Operational quantum average-case distances},\ }\href {https://doi.org/10.22331/q-2023-09-11-1106} {\bibfield  {journal} {\bibinfo  {journal} {Quantum}\ }\textbf {\bibinfo {volume} {7}},\ \bibinfo {pages} {1106} (\bibinfo {year} {2023})}\BibitemShut {NoStop}%
\bibitem [{\citenamefont {Nielsen}\ and\ \citenamefont {Chuang}(2012)}]{Nielsen2012}%
  \BibitemOpen
  \bibfield  {author} {\bibinfo {author} {\bibfnamefont {M.~A.}\ \bibnamefont {Nielsen}}\ and\ \bibinfo {author} {\bibfnamefont {I.~L.}\ \bibnamefont {Chuang}},\ }\href {https://doi.org/10.1017/cbo9780511976667} {\emph {\bibinfo {title} {Quantum Computation and Quantum Information}}}\ (\bibinfo  {publisher} {Cambridge University Press},\ \bibinfo {year} {2012})\BibitemShut {NoStop}%
\end{thebibliography}%

\end{document}